\documentclass[11pt]{amsart}

\usepackage{amssymb}
\usepackage{amsmath}
\usepackage{amsbsy}
\usepackage{amsthm}
\usepackage{amscd}
\usepackage{amsfonts}
\usepackage{graphicx}
\usepackage{subfigure}
\usepackage{color}
\usepackage{comment}

\usepackage{epsfig,amsfonts}
\usepackage{graphicx,epstopdf}
\usepackage{psfrag}
\usepackage{array}
\usepackage{color,soul}
\soulregister\cite{7}
\soulregister\citep{7}
\soulregister\ref{7}
\soulregister\pageref{7}
\usepackage[usenames,dvipsnames]{xcolor}
\sethlcolor{yellow}
\usepackage{appendix}

\usepackage{enumerate}
\usepackage{lineno}
\usepackage{placeins}
\usepackage{booktabs,ctable,multirow,longtable}
\usepackage{changepage}

\usepackage[linesnumbered,ruled]{algorithm2e}

\SetCommentSty{mycommfont}
\usepackage{setspace}

\DeclareMathAlphabet{\mathpzc}{OT1}{pzc}{m}{it}
\DeclareMathAlphabet{\mathcalligra}{OT1}{calligra}{m}{it}
\newcommand{\eqn}[2]{\begin{equation} \label{#1} {#2} \end{equation}}

\newcommand{\bs}[1]{\boldsymbol{#1}}

\DeclareMathOperator*{\argmin}{arg\,min}
\newtheorem{thm}{Theorem}[section]

\theoremstyle{definition}

\theoremstyle{remark}

\theoremstyle{problem}
\newtheorem{problem}[thm]{Problem}

\graphicspath {
    {./}
    {./Figures/}
    {./Images/}
    {./Images/EPS/}
}

\begin{document}

\title[Data-Driven Rate-Dependent Fracture Mechanics]{Data-Driven Rate-Dependent Fracture Mechanics}

\author{P.~Carrara$^{*}$}
\address[P.~Carrara$^{*}$]{Dept. of Mechanical and Process Engineering, ETH Z\"urich, Tannenstr. 3, 8092 Zurich, Switzerland }
\email{pcarrara@ethz.ch}
\thanks{$^{*}$\textit{Corresponding author} pcarrara@ethz.ch}

\author{M.~Ortiz}
\address[M.~Ortiz]{Division of Engineering and Applied Science, California Institute of Technology, 1200 East California Boulevard, Pasadena, CA 91125, USA.}
\email{ortiz@caltech.edu}

\author{L.~De Lorenzis}
\address[L.~De Lorenzis]{Dept. of Mechanical and Process Engineering, ETH Z\"urich, Tannenstr. 3, 8092 Zurich, Switzerland }
\email{ldelorenzis@ethz.ch}


\begin{abstract}
 We extend the model-free data-driven paradigm for rate-independent fracture mechanics proposed in \textit{Carrara et al. (2020), Data-driven Fracture Mechanics, Comp. Meth. App. Mech. Eng., \textbf{372}}  to rate-dependent fracture and sub-critical fatigue. The problem is formulated by combining the balance governing equations stemming from variational principles with a set of data points that encodes the fracture constitutive behavior of the material. The solution is found as the data point that best satisfies the meta-stability condition as given by the variational procedure and following a distance minimization approach based on closest-point-projection. The approach is tested on different setups adopting different types of rate-dependent fracture and fatigue models affected or not by white noise.
\end{abstract}

\maketitle

\tableofcontents
%\linenumbers

\section{Introduction}

	Model-free data-driven computational mechanics has been recently proposed in \cite{Kirchdoerfer2016,Conti2018}. It aims at retaining the epistemic and certain conservation laws governing a mechanics problem, while avoiding the introduction of uncertain constitutive relationships. The uncertainty of the latter stems from the attempt to distillate a postulated and empirical constitutive relation from a set of material observations \cite{Lopez2018,Carrara2020}. This introduces approximations, heuristics and an implicit interpretation of the material data mediated by the specific form of the relationship  chosen, introducing a subjective bias \cite{Carrara2020}. While in the past analytical constitutive models have been used to generalize few observations, today this concept can be overcome since the modern world is data-rich. Hence, the main idea is to introduce the raw material data directly into the solution stream of the mechanics boundary value problem \cite{Kirchdoerfer2016,Kirchdoerfer2017,Kirchdoerfer2018,Ibanez2017}. 

	In general, a model-free data-driven solver seeks, within the material data set, the point to be associated to a specific state that best fulfills the basic epistemic conservation laws \cite{Kirchdoerfer2016,Conti2018}. The identification of the solution is done by minimizing the distance induced by a metric defined in the so-called phase-space between the material data set and the subspace of states that are compatible and in equilibrium. Further developments of this basic approach encompass a maximum entropy strategy to increase the robustness with respect to noisy material data sets \cite{Kirchdoerfer2017}, the reformulation of the problem as a mixed-integer optimization scheme \cite{Kanno2019} and the introduction of local manifold learning schemes that tessellate the material space with linear embeddings \cite{Eggersmann2020,Kanno2020}. Most of the numerical studies available in the literature deal with conservative systems, namely under the assumption of linear elasticity \cite{Kirchdoerfer2016,Kirchdoerfer2017,Conti2018}, geometrically non-linear elasticity \cite{Conti2020,Nguyen2018} and elastodynamics \cite{Kirchdoerfer2018}.  

Data-driven computational mechanics can be also adopted in the context of inverse problems such as in \cite{Leygue2018,Stainier2019} where distance minimization is used to identify material parameters from experimental tests. Also, in \cite{Ibanez2017,Lopez2018} a machine learning technique is devised able to identify numerical constitutive manifolds from experimental data, while in \cite{Flaschel2020} interpretable hyperelastic constitutive models are discovered from synthetic data sets.
	
	The extension of the data-driven paradigm to dissipative materials is, to date, a largely open question \cite{Carrara2020}. Here, the main issue is how to ensure data representability when the material behavior becomes history dependent without introducing modeling hypotheses such as \textit{a priori} defined dependencies from postulated internal/history variables. In this context, Eggersmann et. al \cite{Eggersmann2019} consider specific representational paradigms governing the evolution of the material data set, while in Ladeveze et al. \cite{Ladeveze2019} the solution of the data-driven problem is conditioned by the accumulated plastic strain-rate. In Carrara et al. \cite{Carrara2020}, a model-free data-driven approach to variational fracture mechanics is proposed. Here, the natural choice for the history variable is the crack length, which is an experimentally measurable quantity. Solutions stemming from both local and global stability principles are compared and different distances to be minimized are defined. 
	
	In \cite{Carrara2020} the analysis is limited to quasi-static rate-independent conditions. Rate-dependent crack propagation is typical of many common materials, such as soft materials for biomechanics applications and composite materials for the civil and aerospace industry \cite{Lefranc2014,Hauch1998}. Moreover, although being exhaustive for smooth evolutions, the rate-independent  framework is not representative of discontinuous processes, namely in case of crack jumps. Negri \cite{Negri2010b} demonstrated that, while for smooth crack evolutions the rate-independent quasi-static problem coincides with the rate-dependent one for vanishing loading rates, this no longer hold when dealing with discontinuous evolutions. This stems from the fact that the discontinuities in the crack propagation evolution can be physically understood as an indirect manifestation of dynamic processes. Another issue related to  discontinuous crack evolution is the possible competition between different locally stable states occurring in case of non-convex free-energy potentials \cite{Negri2010, Negri2010b, Carrara2020}. Within the classical analytical solution process, one can select the correct solution invoking the causality or Onsager's principle \cite{Negri2010} but in the model-free data-driven paradigm such considerations cannot play any role. As illustrated in \cite{Carrara2020}, this might lead the data-driven search procedure to favor the solution associated with the maximum dissipation, regardless of whether initial and final states are separated by energetic barriers. This issue is prevented in a rate-dependent framework thanks to viscous regularizing effects \cite{Negri2010b,Toader2009,Knees2008}. Another limitation of the rate-independent framework is that it inhibits crack propagation below a certain energetic threshold, which excludes the occurrence of sub-critical cyclic crack growth. Instead, this can be accounted for in analogy with the rate-dependent framework upon the adoption of some appropriate precautions \cite{Larsen2009}. 
	
	In this paper we extend the model-free data-driven approach proposed in \cite{Carrara2020} to the case of rate-dependent and fatigue fracture propagation. To this end, we adopt the approach of \cite{Negri2010b,Larsen2009}, where a rate-dependent energy dissipation is assumed to account for crack micro-branching occurring at non-negligible crack tip velocities \cite{Fineberg1999,Sharon1999,Lefranc2014,Hauch1998,Ravi2004}. Although including this effect goes in the direction of a dynamic description of the cracking process, it is not directly governed by inertial forces that can thus be neglected \cite{Negri2010b}\footnote{This hypothesis applies as long as the crack tip velocity is much lower than the Rayleigh wave speed \cite{Fineberg1999,Sharon1999,Lefranc2014}.}.
	
	To focus our attention on the data-driven formulation we consider the simplest case of a crack propagating along a known direction in a linear elastic body with known compliance function. Under these assumptions the crack size and propagation velocity are expressed by two scalar quantities. Following \cite{Kirchdoerfer2016, Carrara2020}, we remove any fracture modeling hypothesis and we let the fracture constitutive behavior be embodied by a set of data points. The constraint set encoding the epistemic conservation laws is derived following variational principles and the data-driven solution is found through a local or meta-stability principle \cite{Negri2010b}. Note that, being the rate-dependent problem generally convex, in principle globally and locally stable solutions coincide. However, as illustrated in \cite{Carrara2020}, the adoption of a meta-stability approach mitigates the over-sensitivity of the global minimization approach to the noise possibly present in the data set. Also, the distance minimization procedure is based on the closest-point-projection strategy which was shown in \cite{Carrara2020} to outperforms the other investigated strategies.
	
	The paper is structured as follows. The variational formulation of rate-dependent fracture is summarized along with its adaptation to sub-critical fatigue conditions in Section~\ref{sct:theory}. Section~\ref{sct:DD_formul} illustrates the proposed data-driven procedures along with the details of the numerical implementation. Section~\ref{sct:num_ex} compares standard and data-driven results under both rate-independent and -dependent conditions and for sub-critical fatigue crack growth, adopting material data sets with and without noise. For rate-dependent fracture, cases featuring dependence on the crack tip velocity only or on both crack size and velocity are considered and the ability of the proposed approach to reproduce and extend the rate-independent case is demonstrated. Concluding remarks are drawn in Section~\ref{sct:conclusions}.

\section{Rate-dependent fracture mechanics} \label{sct:theory}

	Aim of this section is to characterize the equilibrium of a cracked body showing a rate-dependent fracture propagation behavior. We also show that the same framework can be adapted to the study of sub-critical fatigue crack growth along the same lines as in \cite{Negri2010b,Larsen2009}. Inertial forces are assumed to be negligible and the microscopic rate-dependent effects are lumped in the dissipation potential. This allows to study crack propagation under static equilibrium conditions, where, however, the energy dissipated by the moving crack tip depends on its velocity.

		\subsection{Preliminary definitions and problem statement}
	
	We assume a straight and planar crack propagating under pure mode-I conditions in an otherwise linear elastic body. Under these conditions the load is described by an effective force $P$, the deformation by an effective conjugate displacement $\Delta$ and the crack extension by its length $a$, which is the result of the propagation history described by the (scalar) crack tip velocity $v$. The relation between $a$ and $v$ reads
 \eqn{eq:vel}{\dot a(t) =\frac{da(t)}{dt}= v\left(t\right)\ge0\,.}	
 \noindent  where $t$ is the time.  We postulate $v\ge 0$ to enforce crack irreversibility. The crack length at a given instant $t$ is obtained as
	\eqn{eq:prop_eq}{a\left(a_0,\,v,\,[0,\,t]\right) = a_0 + \int_0^t v\left(s\right)\, ds\,.}
\noindent where $a_0$ is the initial crack extension, while we assume without loss of generality that $v_0=v(0)=0$. Note that the crack length at a given instant $t$ depends on the whole crack tip velocity history $v\in[0,\,t]$.

	To simplify the notation, the evolving quantities are endowed with a subscript indicating the time variable, for instance $v(t) = v_t$, $a(a_0,\,v,\,[0,\,t])=a_t$, $\Delta(t) = \Delta_t$ and $P(t)=P_t$.

	Assuming now a displacement-controlled process, the time-continuous evolution problem can be formulated as
 	\begin{problem}[Time-continuous evolution]\label{prob:time_evol}
	Given an initial crack length $a_0$ and the imposed displacement history $\Delta_t$, determine the evolution of the crack tip velocity $v_t$, the crack length $a_t$ and the load $P_t$.
	\end{problem}

	\subsection{Incremental loading procedure}\label{sct:discr_form}
	
	When dealing with an incremental loading procedure it is more convenient to formulate the evolution problem in terms of the finite crack length increment $\bs{\Delta} a$\footnote{The bold-face symbol $\boldsymbol{\Delta}\bullet$ stands here for a range of values or an increment of the quantity $\bullet$ and not for a displacement that is conversely indicated with a light-face $\Delta$.}. Hence, Problem~\ref{prob:time_evol} can be reformulated as
	\begin{problem}[Discrete evolution]\label{prob:discr_evol}
	Let $\bs{\Delta} t\ge0$ be a time interval and $t_k = k\, \bs{\Delta} t$ be a uniform time discretization with  $k$ the (positive integer) time step. At the load step $k+1$, given $v_k$ and $a_k$ at time $t_k$ and $\Delta_{k+1}$ at time $t_{k+1}$, determine the crack tip velocity $v_{k+1}\ge0$, the (finite) increment of the crack length $\bs{\Delta} a_{k+1}$ so that $a_{k+1}=a_k+\bs{\Delta} a_{k+1}$, and the load $P_{k+1}$.
	\end{problem}

\noindent In this case a time integration strategy should be adopted to obtain the crack size increment $\bs{\Delta} a_{k+1}$. This point is particularly relevant for the numerical solution of the problem; it is illustrated in detail in the following sect.~\ref{sct:num_proc}.

	\subsection{Energetic quantities and variational approach}\label{sct:var_form}
	
	The evolution problem can be formulated variationally. Let us first introduce the relevant energetic quantities. The free energy of the system is written as
	\eqn{eq:free_en}{F(\Delta,\,a) = E(\Delta, a) + F_R(a, [0,\,t])\quad  \text{with}\ a\ \text{given by}\ (\ref{eq:prop_eq})\,,}
\noindent where $E(\Delta,\,a)$ is the elastic strain energy and $F_R(a, [0,\,t])$ is the energy dissipated to create new fracture surfaces. The latter is defined starting from a dissipation potential $\mathcal{D}(a,v)$ such that
 	\eqn{eq:res_term}{F_R(a,\,[0,\,t]) = \int_0^t \mathcal{D}(a_s,v_s)\,ds\,.}
\noindent Note that the dissipation potential depends on the current values of crack size and tip velocity, while the resistance term $F_R$ is a function of the whole crack evolution history.  Also, the energy dissipated per unit crack length or \emph{critical energy release rate} $G_R(a,v)$ is defined starting from the dissipation potential as \cite{Negri2010b}
	\eqn{eq:res_ERR}{G_R(a,v)=\frac{dF_R}{da}(a,\,[0,\,t])=\frac{1}{v}\frac{dF_R}{dt}(a,\,[0,\,t])=\frac{\mathcal{D}}{v}(a,\,v) \,,}
	\noindent where (\ref{eq:vel}) is used.

	Also, inspired by a large number of experimental studies, the following properties are usually assumed to hold
	\eqn{eq:RI_ERR}{\begin{split} \lim_{v\to0}G_R(a,v) = G_R^{QS}(a),\ \  \lim_{v\to+\infty}G_R(a,v)=+\infty\ \ \text{and}\ \ \frac{\partial G_R}{\partial v}(a,v)\ge0\\ \forall a\ge a_0\,,\end{split}}
\noindent where $G_R^{QS}(a)$ is the critical quasi-static rate-independent energy release rate, which reduces to the Griffith critical energy release rate if $G_R^{QS}(a)=G_c=$ constant. Evidently, if the resistance energy release rate function is independent from the crack length, i. e. $G_R(a,v)=G_R(v)$, the property (\ref{eq:RI_ERR}a) delivers an implicit Griffith-like quasi-static resistance model.  It is also worth noting here that (\ref{eq:RI_ERR}c) renders the problem associated with the rate-dependent fracture propagation convex \cite{Negri2010b}, so that global and local minimization approaches became equivalent \cite{Carrara2020}. 
	
	The elastic strain energy $E$ is written in term of the compliance function as follows
	\eqn{eq:el_en}{E(\Delta,a) = \frac{\Delta^2}{2C(a)}\,,}
\noindent where $C(a)$ is the (known) \textit{compliance function} of the body. The elastic strain energy has the following properties
	\eqn{eq:el_en_prop}{P = \frac{\partial E}{\partial \Delta}(\Delta,a)=\frac{\Delta}{C(a)}\ \ \text{and}\ \ G=-\frac{\partial E}{\partial a}(\Delta,a)=\frac{\Delta^2}{2C^2(a)}\frac{dC}{da}\,,}
	\noindent where $G$ is termed energy release rate. 
	
	The solution to Problem~\ref{prob:time_evol} is based on the minimization of the free energy (\ref{eq:res_term}). As in \cite{Negri2010b}, we consider the time-discrete Problem~\ref{prob:discr_evol} where the crack tip velocity at the step $k+1$ is approximated as
	\eqn{eq:incr_v}{\bar v_{k+1}=\frac{a_{k+1}-a_k}{\bs{\Delta} t}\,.}
\noindent Also the dissipated energy is decomposed as follows
	\eqn{eq:diss_decomp}{\begin{split}F_R(a_{k+1},\,[0,\,t_{k+1}])&=F_R(a_{k},\,[0,\,t_{k}])+\bs{\Delta}F_R(a_{k+1},\,[t_k,\,t_{k+1}])\\ &=F_R(a_{k},\,[0,\,t_{k}])+\int_{t_k}^{t_{k+1}} \mathcal{D}(a_s,v_s)\,ds\,.\end{split}}
	\noindent The free energy $ F(\Delta_{k+1},a_{k+1})$ at time $t_{k+1}$ can be approximated as
	\eqn{eq:incre_en}{\begin{split}F(\Delta_{k+1},a_{k+1})\,\simeq\ &\tilde F(\Delta_{k+1},a_{k+1}) = E(\Delta_{k+1},a_{k+1})+\\& + F_R(a_{k},\,[0,\,t_{k}])+\underbrace{\mathcal{D}\left(a_{k+1},\,\frac{a_{k+1}-a_k}{\bs{\Delta} t}\right)\bs{\Delta}t}_{\simeq \bs{\Delta}F_R(a_{k+1},\,[t_k,\,t_{k+1}])}\,.\end{split}}

	The solution $a_{k+1}^*$ given the load $\Delta_{k+1}$ and the solution at the previous time step $a^*_{k}$ is then found as
	\eqn{eq:min}{a^*_{k+1} \rightarrow\argmin_{a_{k+1}\ge a_k^*}\left\{ \tilde F(\Delta_{k+1},a_{k+1})\right\}\,.}
	\noindent Adopting a local minimization strategy, we aim thus at finding the crack size $a_{k+1}^*$ such that
\eqn{eq:locmin}{\begin{split} \exists h>0\ :\ \tilde F(\Delta_{k+1},a_{k+1}^*)&\le \tilde F(\Delta_{k+1},a_{k+1}^*+\delta a)\,,\\\forall\ &|\delta a|\le h\ \text{such that}\ a_{k+1}^*+\delta a \ge a_k^*\,.\end{split} }
\noindent Expanding (\ref{eq:locmin}) in Taylor series up to the first order we can write
	\eqn{eq:first_ord}{\exists h>0 \ :\ \left.\frac{\partial \tilde F}{\partial a}\right|_{a_{k+1}^*}\hspace{-3mm}\delta a\ge 0\,, \quad \forall\ |\delta a|\le h\ \text{such that}\ a_{k+1}^*+\delta a \ge a_k^*\,.}

 Introducing (\ref{eq:res_ERR}) and (\ref{eq:el_en_prop}b)  we obtain
\eqn{eq:min_4}{\begin{split}\left[-G(\Delta_{k+1},\,a_{k+1}^*)+G_R\left(a_{k+1}^*,\,\frac{a_{k+1}^*-a_k^*}{\bs{\Delta} t}\right)\right]&\delta a\ge 0\,, \\ \forall\ |\delta a|\le h\ \text{such that}&\ a_{k+1}^*+\delta a \ge a_k^*\,.\end{split}}

The fulfillment of (\ref{eq:min_4}) gives the following activation and propagation criteria
\eqn{eq:activation_discr}{\begin{split}\text{if}\ a_{k+1}^*=a_k^*, \quad (\bar v_{k+1}^*=0) \ \Rightarrow \ \delta a& \ge 0\ \ \text{then}\\ &G(\Delta_{k+1}, a_{k+1}^*) -G_R^{QS}\left(a_{k+1}^*\right) \le 0\,;\end{split}}

\eqn{eq:propagation_discr}{\begin{split}\text{if}\ a_{k+1}^*>a_k^*, \quad (\bar v_{k+1}^*>0)\ \Rightarrow \ \ &\delta a \lesseqgtr 0\ \ \text{then}\\ G(\Delta_{k+1},& a_{k+1}^*) -G_R\left(a_{k+1}^*,\,\frac{a_{k+1}^*-a_k^*}{\bs{\Delta} t}\right) = 0\,.\end{split}}
\noindent The conditions (\ref{eq:activation_discr})-(\ref{eq:propagation_discr}) define by induction the series $\tilde a_{\bs{\Delta}t} =\{a_k^*,\ k = 1,...,N\}$ and $\tilde v_{\bs{\Delta}t} =\{\bar v_k\ \text{given by}\ (\ref{eq:incr_v}),\ k = 1,...,N\}$ as the solution of the incremental minimization problem (\ref{eq:min}) and, hence, of Problem~\ref{prob:discr_evol}. As proven in \cite{Negri2010b}, the series minimizing (\ref{eq:incre_en}) converge uniformly to the solution of Problem~\ref{prob:time_evol} when $\bs{\Delta}t\to 0$, namely $\tilde a_{\bs{\Delta}t} \to a_t$ and $\tilde v_{\bs{\Delta}t} \to v_t$. In the time-continuous case (\ref{eq:activation_discr})-(\ref{eq:propagation_discr}) can be rewritten as
\eqn{eq:activation}{\text{if}\ v_t=0 \ \Rightarrow \ G(\Delta_t, a_t) -G_R^{QS}\left(a_t\right) \le 0\,;}

\eqn{eq:propagation}{\text{if}\  v_t>0\ \Rightarrow \  G(\Delta_t, a_t) -G_R\left(a_t,\,v_t\right) = 0\,.}

	The criteria (\ref{eq:activation})-(\ref{eq:propagation}) can be recast as a set of Kuhn-Tucker conditions as follows
\eqn{eq:KT_cond}{\begin{aligned} v_t &\ge 0\,, \\ G(\Delta_t, a_t) -G_R(a_t,v_t) &\le 0\,, \\ \left[ G(\Delta_t, a_t) -G_R(a_t,v_t)\right]v_t &= 0\,.\end{aligned}}
\noindent Under some regularity conditions, (\ref{eq:KT_cond}) can be converted to the following ordinary differential equation \cite{Negri2010b}
\eqn{eq:v_ODE}{ v_t = \frac{da_t }{dt}= \tilde G_R^{-1}\left(G(\Delta_t,a_t)\right)\,,}
\noindent where $\tilde G_R^{-1}\left(G(\Delta_t,a_t)\right)$ is defined as
\eqn{eq:ODE2}{\begin{split}\tilde G_R^{-1}&\left(G(\Delta_t,a_t)\right) = \\[5pt] =&\begin{cases} G_R^{-1}\left(G(\Delta_t,a_t)\right)\,, & \text{if}\ G(\Delta_t,a_t) > G_R^{QS}(a_t)\\
0\,, & \text{otherwise}\,.\end{cases}\end{split}}

	\subsection{Body connected to a loading device}

	In case the body is loaded through a simple linear device of known compliance $C_M$ and the test is driven controlling the device displacement $\Delta_T$, the opening displacement $\Delta$ and the load $P$ are related by 
	\eqn{eq:device_rel}{\Delta_T = \Delta+ C_MP\,.}
\noindent Substituting (\ref{eq:el_en_prop}a) into (\ref{eq:device_rel}) provides the following relation between crack opening and device displacement
	\eqn{eq:disp_machine}{\Delta=\frac{C(a)}{C(a)+C_M}\Delta_T\,.}

  In this case, the variational procedure requires the definition of the following function \cite{Carrara2020}
	\eqn{eq:var_fct}{\Phi(\Delta, P, a) =E(\Delta,a) + F_R(a,\,[0,\,t]) - \frac{C_M}{2}P^2-P(\Delta-\Delta_T)\,.}
\noindent Combining (\ref{eq:el_en}), (\ref{eq:el_en_prop}a) and (\ref{eq:disp_machine}) into (\ref{eq:var_fct}) permits to eliminate $P$ and $\Delta$ and gives the following modified free energy function
	\eqn{eq:free_en_device}{F_M(\Delta_T,a) = E_M(\Delta_T, a) + F_R(a, [0,\,t])\,,}
	
	\noindent with
	\eqn{eq:el_en_device}{E_M(\Delta_T, a) =\frac{1}{2}\frac{\Delta_T^2}{C(a)+C_M}\quad \text{and}\quad G(\Delta_T,a)=\frac{\Delta_T^2}{2(C(a)+C_M)^2}\frac{dC}{da}\,.}
	
	Imposing a load ramp $\Delta_{T,\,t}$ the energy (\ref{eq:free_en_device}) can be minimized as in sect.~\ref{sct:var_form}, ultimately giving activation and propagation criteria similar to (\ref{eq:activation})-(\ref{eq:propagation}). In this case, the loading history $\Delta_t$ should be replaced with $\Delta_{T,\,t}$ and the energy release rate with (\ref{eq:el_en_device}b).

	\subsection{Sub-critical crack growth and fatigue}\label{sct:fatigue}
	
	The activation criterion (\ref{eq:activation}) states that when the energy release rate at the crack tip is below the critical quasi-static rate-independent limit $G_R^{QS}(a)$ the crack propagation is inhibited. However, the experimental evidence shows that the crack might still be able to propagate even below $G_R^{QS}(a)$ but its growth rate is so low that it becomes measurable only after the application of several load cycles. 
	
	This observation allows us to describe the cyclic or fatigue behavior within the same framework presented in sects.~\ref{sct:var_form}-\ref{sct:discr_form}. In this case, the role of time is played by the number of load cycles $N$ and the sub-critical crack tip velocity $v_N$ must be intended as
		\eqn{eq:sub_crit_vel}{v_N=\frac{da}{dN}\,,} 
\noindent and takes the name of sub-critical or fatigue crack growth rate. Concerning the resistance curve, the experimental evidence suggests that the front growth rate is related to the energy release rate through a power law, possibly with a (fatigue) driving force threshold. The first and most widespread relation of this type is proposed by Paris and Erdogan in \cite{Paris1963} and, since then, has evolved until reaching today the form of the well-known \emph{NASGRO} equation \cite{Rabold2013}. This type of fatigue constitutive laws usually give the crack growth rate as a function of $\boldsymbol{\Delta} K=K_{max}-K_{min}$, where $K_{min}$ and $K_{max}$ are the maximum and minimum values of the stress intensity factor reached in one cycle. The latter quantities can be converted, for mode-I plane states, into energy release rates by means of the relationship $G=K/Y^{\prime 2}$, where $Y^\prime$ is the Young's modulus under either plane stress or strain state, giving
	\eqn{eq:v_fatigue}{v_N=\varphi( \bs{\Delta} G )\quad \text{with } \bs{\Delta}G=\frac{\bs{\Delta}K}{Y^\prime} \,,}
\noindent where $\varphi$ is a known function dependent on the chosen model. Considering that the fatigue behavior in brittle materials depends only on the range of load applied at the crack tip rather than on its absolute value, the energy release range in (\ref{eq:v_fatigue}) can be replaced by an equivalent absolute value. Thus, once the applied energy release rate is specified using, e.~g., (\ref{eq:el_en_prop}b), equation (\ref{eq:v_fatigue}) can be considered the sub-critical version of (\ref{eq:v_ODE}) and its inverse can be understood, similarly to (\ref{eq:res_ERR}), as an equivalent sub-critical energy release rate rate curve $G_f(v_N)$\footnote{Although here the energy release rate resistance curve is given as a function of $v_N$ only, the proposed approach is general and can also encompass the case of dependence upon both $v_N$ and $a$. However, the latter case does not find any support in the literature, where, typically, the fatigue constitutive laws are given in the form (\ref{eq:v_fatigue}).}, i. e.
	\eqn{eq:res_fatigue}{G_f(v_N) = \varphi^{-1}(v_N)\,.}
	
	A little care has to be paid when dealing with Paris-Erdogan-type laws, whose aim is a phenomenological description of the process. In this case the pseudo-time variable $N$ takes only positive integer values and is thus discrete. Also, these laws give the average crack tip advancement within one cycle when a certain nominal $\bs{\Delta} K$ or $\bs{\Delta} G$ is applied at the crack tip. Hence, the evolution of the system must be quantized cycle-by-cycle and the applied range of stress intensity factor or energy release rate must be computed at the crack tip at the beginning of each cycle. In orther words, the evolution of the system within a single cycle must be neglected because it is lumped in the definition of the law.

	In this case we only have a discrete evolution problem, which can be formulated as follows
		\begin{problem}[Sub-critical crack growth]\label{prob:fatigue_prob}
	For cycle $k+1$, given the crack length $a_k$ at the end of cycle $k$ and the applied range of energy release rate $\bs{\Delta} G_{k+1}$, determine the crack growth rate $v_{k+1}\ge0$ and the crack length at the end of the cycle $a_{k+1}=a_k+v_{k+1}\cdot 1\text{ cycle}$.
	\end{problem}
	
Note that, since the fatigue crack propagation process  is intended here as a sequence of rate-independent steps, Problem~\ref{prob:fatigue_prob} does not need the definition of an integration scheme, unlike Problem~\ref{prob:discr_evol}. 

\section{Data-driven approach}\label{sct:DD_formul}

	\subsection{Data representation}
	
	The first problem to face is how to characterize a rate-dependent or cyclic fracture process by means of data. To this end we first revisit the classical solution of the evolution problem, where the functions characterizing the elastic energy $E(\Delta,a)$ and the dissipation potential $\mathcal{D}(a,v)$ are completely known.
	
	Similarly to \cite{Carrara2020}, one possibility is to define the solution of the propagation problem illustrated in sect.~\ref{sct:theory} in terms of pairs $(\Delta_t, P_t)$ lying in the corresponding \emph{phase space} $\mathcal{Z}$ that are both in the \emph{constraint set} defined by (\ref{eq:device_rel}) and in the \emph{material data set}. The latter is defined as the set of points in $\mathcal{Z}$ that, given $a_0$, satisfies simultaneously (\ref{eq:prop_eq}), (\ref{eq:el_en_prop}a) and (\ref{eq:v_ODE}) at every instant $t$. However, this might results in a non-trivial task because of the history dependence of the crack length.

	Rather, we consider the solution being represented by the crack size $a_t$ and corresponding tip velocity $v_t$ that minimizes the free energy (\ref{eq:free_en}) or (\ref{eq:free_en_device}). Taking as paradigmatic example the case of a specimen connected to a loading device, given $\Delta_{T,\,t}$ and $a_0$, the solution of the ordinary differential equation (\ref{eq:v_ODE}) delivers the crack evolution $a_t$ along with the related crack tip velocity $v_t$ that at every instant $t$ fulfill (\ref{eq:KT_cond}). Note that (\ref{eq:KT_cond}) identifies, in case of crack propagation, the state corresponding to the intersection between the functions $G(\Delta_t,\,a_t)$ and $G_R(a_t,\,v_t)$ as solution.
 
	We remove now the hypothesis that the resistance quantities are analytically known, rather they are introduced as a discrete resistance data set $\mathcal{D}_R$. Also, we retain the assumption that the elastic strain energy is exactly known and characterized in terms of the compliance function of the specimen $C(a)$. Within the experimental practice it is customary to provide the resistance quantities as the critical energy release rate $G_R$. Under these premises, (\ref{eq:min}) entails, at every instant $t$, a discrete minimization over the points included into $\mathcal{D}_R$. 

	\subsection{Data-driven computational procedure}\label{sct:num_proc}
		
	Although the data-driven procedure that we propose is unique and based on closest-point projection proposed in \cite{Carrara2020}, as follows we distinguish between different cases.

		\subsubsection{Implicit quasi-static Griffith model} \label{sct:Griff}
		Many experimental results in the literature involve resistance data that are independent on the crack length \cite{Lefranc2014,Hauch1998,Fineberg1999,Sharon1999}. Hence, $\mathcal{D}_R$ is composed of pairs $(\hat v_i,\,\hat G_{R,i})$, implicitly involving a Griffith-like rate-independent limit.
		
	In this case, we first define the quasi-static rate-independent critical energy release rate $\hat G_R^{QS}$ as the value $\hat G_{R,i}$ corresponding to $\hat v_i=0$.  This value remains constant throughout the computation for any value of crack size $a$, which conversely varies assuming thus the role of a history variable. Then, the activation criterion (\ref{eq:activation}) at every instant $t$ must be verified and, if the condition for the crack propagation is satisfied, the crack tip velocity  must be defined so as to best approximate (\ref{eq:propagation}), namely $v_t=\hat v_{i^*_t}$ where
\eqn{eq:i_min}{i^*_t=\argmin_{i}\left\{G(\Delta_{T,\,t},a(v_t))-\hat G_{R,i}\right\}\,,}
\noindent where the term into brackets can be regarded as a generalized distance between the constraint set and the material data set.

	As mentioned in sect.~\ref{sct:discr_form}, within an incremental loading process a suitable time integration procedure is needed and here we adopt a Crank-Nicholson implicit scheme. Hence, with $v_k$ and $v_{k+1}$ denoting respectively the initial and final (unknown) crack tip velocities, the crack length increment can be approximated as
	\eqn{eq:int_scheme}{\bs{\Delta} a_{k+1} = \frac{v_k + v_{k+1}}{2}\bs{\Delta} t\,. }	
\noindent We adopt a solution procedure using a predictor-corrector scheme. At the load step $k+1$, given  $a_k$ and $v_k$ at $t_k$ along with $\Delta_{T,k+1}$ at $t_{k+1}$,  we define the trial crack length increment as
\eqn{eq:trial_da}{\bs{\Delta} a^{TRIAL}_{k+1}=\frac{v_k}{2}\bs{\Delta} t\,.}
Then, we check if $G(\Delta_{T,k+1},a_k+\bs{\Delta} a^{TRIAL}_{k+1})-\hat G_R^{QS}\le0$. If this is the case, then the solution $\bs{\Delta} a_{k+1}=\bs{\Delta} a^{TRIAL}_{k+1}$ and the crack at $t_{k+1}$ is at equilibrium, namely $v_{k+1}=0$. Conversely if $G(\Delta_{T,k+1},a_k+\bs{\Delta} a^{TRIAL}_{k+1})-G_R^{QS}>0$, the evolution of the crack during the load step must comply with (\ref{eq:propagation}). At $t_{k+1}$ we are thus looking for the best discrete approximation of $G(\Delta_{T,k+1},a_k+\bs{\Delta} a_{k+1})-G_R(v_{k+1})=0$. However, unlike in the time-continuous evolution, now the evolution of the crack length within the finite time increment must be accounted for.

To address this point, for each pair $(\hat v_i,\,\hat G_{R,i})\in\mathcal{D}_{R}$ we compute the distance
	\eqn{eq:dist}{\begin{split}&d_{i,\,k+1}=\min_{ a \ge a_k}\left\{\vphantom{\sqrt{\left(\hat G_{R,i}\right)}}\right. \\&\left.\sqrt{\left( \displaystyle\frac{v_k + \hat v_{i}}{2}\bs{\Delta} t+a_k- a\right)^2+\left(\hat G_{R,i} - G\left(\Delta_{T,k+1}, a\right)\right)^2}\vphantom{\sqrt{\left(\hat G_{R,i}\right)}}\right\}\,,\end{split}}
\noindent and we find the solution as $v_{k+1}=\hat v_{i^*_{k+1}}$, with 
	\eqn{eq:i_min_incr}{i^*_{k+1}=\argmin_{i}\left\{d_{i,\,k+1}\ :\ (\hat v_{i},\,\hat G_{R,i})\in\mathcal{D}_{R}\right\}\,.}

		\subsubsection{Complete model} \label{sct:3D}
		In the most general case the resistance dataset may be composed of triplets $(\hat v_i,\,\hat a_i, \,\hat G_{R,i})$ and hence the closest-point-projection must take place in the 3D space $(v,\,a,\,G)$. In this case, the quasi-static rate-independent critical energy release rate is no longer a constant parameter as in sect.~\ref{sct:Griff}, rather it can be defined as a subset $\mathcal{D}_R^{QS}\subset\mathcal{D}_R$ of points characterized by a crack tip velocity $\hat v_i$ below a certain threshold related to the instrumental accuracy. Differently than the implicit Griffith model, here the history variable is vectorial and constituted by the couple $(a_t,\,\hat G_{R,\,t}^{QS})$. 
		
		For the time-continuous evolution, we define the current value of the quasi-static critical energy release rate $\hat G_{R,\,t}^{QS}$ as the value $\hat G_{R,i}$ corresponding to $\hat a_i = a_t$. Then, we proceed  to check the activation criterion (\ref{eq:activation}) and, if the crack fulfills the propagation condition, the solution is found as $v_t=\hat v_{i^*}$ with
	\eqn{eq:i_min_3D_1}{\begin{split}&i^*_t=\argmin_{i}\left\{\vphantom{\sqrt{\left(\hat a_{i}- a(v_t)\right)^2+\left(\hat G_{R,i} - G(\Delta_{T,\,t}, a(v_t)\right)^2}}\right. \\ &\left. \sqrt{\left(\hat a_{i}- a_t \right)^2+\left(\hat G_{R,i} - G(\Delta_{T,\,t}, a_t\right)^2}\, :\, \hat a_i \ge a_t\right\}\,.\end{split}}

\noindent Again, the term into brackets can be regarded as a generalized distance.

		Compared to the implicit Griffith case, the incremental loading case requires a little more care. At the step $k+1$, given  $a_k$ and $v_k$ at $t_k$ along with $\Delta_{T,k+1}$ at $t_{k+1}$, we first determine the trial crack length increment as in (\ref{eq:trial_da})  and $\hat G_{R,k+1}^{QS}$ as the value $\hat G_{R,i}\in\mathcal{D}_R^{QS}$ corresponding to $\hat a_i = a_k+\bs{\Delta} a^{TRIAL}_{k+1}$. Then, if $G(\Delta_{T,k+1},a_k+\bs{\Delta} a^{TRIAL}_{k+1})-G_{R,k+1}^{QS}\le0$ we have $\bs{\Delta} a_{k+1}=\bs{\Delta} a^{TRIAL}_{k+1}$ and $v_{k+1}=0$. Conversely, if $G(\Delta_{T,k+1},a_k+\bs{\Delta} a^{TRIAL}_{k+1})-G_{R,k+1}^{QS}>0$, to be able to compute the projection of the points in $\mathcal{D}_{R}$ onto the energy release rate function, we parametrize the latter as 
		\eqn{eq:ERR_par}{\begin{cases}G=G(\Delta_{T,k+1},a(v))\,,\\[5pt]
												a(v) =a_k+\displaystyle \frac{v_k+v}{2}\bs{\Delta} t\,, \\[5pt]
												v \,.\\
		\end{cases}}
\noindent which corresponds to projecting $G(\Delta_{T,k+1},a)$ onto the plane $\left(G,\,a(v)\right)$. We can thus define for each triplet $(\hat v_i,\,\hat a_i, \,\hat G_{R,i})\in \mathcal{D}_R$, the distance
		\eqn{eq:dist_3D}{\begin{split}&d_{i,\,k+1}=\min_{v \ge 0}\left\{\ \vphantom{\sqrt{\left(\hat G_{R,i}\right)}}\right.\\ 
&\left. \sqrt{\left(\hat v_{i}-v\right)^2+\left(\hat a_i- a(v)\right)^2+\left(\hat G_{R,i} - G(\Delta_{T,k+1}, a(v)\right)^2}\right\}\,,\end{split}}
\noindent and find the solution as $v_{k+1}=\hat v_{i^*}$, with 
	\eqn{eq:i_min_incr}{i^*_{k+1}=\argmin_{i}\left\{d_{i,\,k+1}\ :\ ( \hat v_i,\,\hat a_i,\,\hat G_{R,i})\in\mathcal{D}_{R},\ \hat a_i\ge a_k\right\}\,.}
	
	\noindent Note that the crack size is here used as a history variable that conditions the choice of the crack tip velocity.

	The advantage in using the complete approach is twofold. On one hand, it makes it possible to reproduce situations where the fracture energy is dependent also on the crack length, as in case of an R-curve rate-independent limit. On the other hand, it might include also the rate-independent case provided that the resistance data set is endowed with observations made in conditions fulfilling the rate-independent assumptions. In this case, the solution is then defined as
	\eqn{eq:3D_sol}{(v_{k+1},\,\bs{\Delta} a_{k+1})=\begin{cases} (0,\,\hat a_{i^*}-a_k) & \text{if }\hat v_{i^*}\in\mathcal{D}_R^{QS} \quad \text{(\textsf{Rate-indep. process})}\\[8pt]
	\left(\hat v_{i^*},\,\displaystyle\frac{v_k+v_{i^*}}{2}\bs{\Delta} t\right)& \text{Otherwise} \quad \text{(\textsf{Rate-dep. process})}\,.
	\end{cases}}
	\noindent The procedure is thus able to recognize when the crack tip velocity is so low that a rate-independent quasi-static state is a good approximation of the solution. In this sense it encompasses and extends the rate-independent approach proposed in \cite{Carrara2020}.

		\subsubsection{Sub-critical crack growth and fatigue} 
		For the sub-critical crack growth, the resistance dataset $\mathcal{D}_{R,f}$ is composed of pairs $(\hat v_{i},\, \hat G_{f,i})$, where $\hat G_{f,i}$ has to be intended as the nominal range of energy release rate spanned during a cycle that triggers a crack growth equal to $\hat v_{i}$ per cycle.
		
		The procedure to solve the sub-critical crack growth problem is very similar to what illustrated in sect.~\ref{sct:Griff} with few fundamental modifications.
		
		Similarly to the implicit quasi-static Griffith model (sect.~\ref{sct:num_Griff}), the limit value for the arrest of the crack propagation $\hat G_f^T$, termed \emph{fatigue threshold}, is a constant parameter. It can be  defined as the value $\hat G_{f,i}$ corresponding to $\hat v_{i}=0$ in $\mathcal{D}_{R,f}$, i.~e. to a negligible crack propagation after a virtually unlimited number of cycles. It is conventionally assumed that a component has an infinite fatigue life if it can survive 2-5$\cdot$10$^6$ cycles.
		
		Consider a specimen with a crack that, at the end of the $k^{th}$ cycle, has a length $a_k$ and during the cycle $k+1$ experiences the application of a cyclic action $\bs{\Delta} \bullet=\bullet_{max}-\bullet_{min}$ where $\bullet$ stands either for $\Delta$, $\Delta_T$ or $P$ in case respectively of a crack opening displacement- (\emph{COD}-), machine displacement- or load-driven test and the subscript $max$ and $min$ refer respectively to the maximum and minimum value reached.  The nominal energy release rate range experienced at the crack tip $D G_{k+1}$ can be thus computed using (\ref{eq:el_en_prop}b) or (\ref{eq:el_en_device}b) as
		\eqn{eq:range_D}{\bs{\Delta} G_{k+1}(\bs{\Delta} \bullet, a_k)=G(\bullet_{max},a_k)- G(\bullet_{min},a_k)\,.}

Considering the nature of the data included into $\mathcal{D}_{R,f}$ (sect.~\ref{sct:fatigue}), the solution is found as $v_N=\hat v_{i^*_{k+1}}$ where
\eqn{eq:i_min_fat}{i^*_{k+1}=\argmin_{i}\left\{\bs{\Delta} G_{k+1}(\bs{\Delta} \bullet, a_k)-\hat G_{f,i}\ :\ \hat G_{R,i}\in\mathcal{D}_{R,f}\right\}\,.}

\noindent In this case, the distance (\ref{eq:dist}) reduces to $\bs{\Delta} G_{k+1}(\bs{\Delta} \bullet, a_k)-\hat G_{f,i}$.

	\subsection{Numerical implementation}\label{sct:num_impl}
	This section provides more details on the algorithms adopted to implement the numerical procedures illustrated in sect.~\ref{sct:num_proc} for the incremental loading procedure.
	
	We first note that the primary variable to solve for is the crack tip velocity $v_{k+1}$, which is always positive, hence, the irreversibility condition is automatically satisfied and encoded into the material data set. The same applies to the properties of the critical energy release rate (\ref{eq:RI_ERR}), that do not have to be enforced. 
	
	For coherence with the irreversibility postulate, we account only for the active part of the energy release rate function while computing the closest-point-projection distance, i.~e. we account only for the portion of $G(\Delta_T, a)$ accessible for $v\ge0$ (or $a \ge a_k$). In the following algorithms, we introduce also the variable $G_{DD,k+1}$ that stores at every load step the energy release rate of the identified solution point.
	
	For all three procedures the computed distance $d_i$ can be seen as a measure of the error committed in fulfilling the basic laws (\ref{eq:KT_cond}). Hence, a tolerance for the distance can be set to reject the solutions that are affected by an excessive error. However, in the following it is assumed that the adopted data set contains enough information to cover the propagation process from the initial crack length $a_0$ to the boundary of the domain.
	
	\subsubsection{Implicit quasi-static Griffith model} \label{sct:num_Griff}
	In Appendix~\ref{app:griffith} the procedure to obtain the data-driven solution in case of implicit quasi-static Griffith model is presented. It might be often the case that no point in the data set corresponds to $\hat v_i=0$, thus the value to be assigned to $\hat G_R^{QS}$ is $\hat G_i\in\mathcal{D}_R$ related to the minimum value of $\hat v_i$.

	\subsubsection{Complete model} \label{sct:num_3D}
	
	The implementation of the complete model is detailed in Appendix~\ref{app:complete}. If no points in $\mathcal{D}_R^{QS}$ corresponds to the initial $a_0$ or to the trial crack length $a_k+\bs{\Delta} a_{k+1}^{TRIAL}$, the initialization and further updates of $\hat G_R^{QS}$ are performed as
	\eqn{eq:avg_GR}{\hat G_{R,k+1}^{QS}=\left\langle \hat G_{R,i_L}^{QS},\,\hat G_{R,i_S}^{QS}\right\rangle	\,,}
	\noindent where $\langle\bullet\rangle$ is the mean value operator and the subscripts $i_L$ and $i_S$ are the indexes of the points in $\mathcal{D}_R^{QS}$ with $\hat a_i$ immediately larger and smaller than $a_0$ or $a_k+\bs{\Delta} a_{k+1}^{TRIAL}$.

	\subsubsection{Sub-critical crack growth and fatigue} \label{sct:num_fatigue}
	The data-driven solution of the sub-critical fatigue crack propagation problem is illustrated in Appendix~\ref{app:fatigue}. The data-driven fatigue threshold $\hat G^T_f$ can be initialized as the value of $\hat G_{f,i} \in \mathcal{D}_{R,f}$ related to the minimum value of $\hat v_{i}$.
	
	Note that in Algorithm~\ref{algo:DD_fat} (Appendix~\ref{app:fatigue}) the parameter $\bs{\Delta} N$ is introduced multiplying the obtained crack growth rate at each step. By setting $\bs{\Delta} N=1$ the effect of each cycle on the crack size is accounted for, obtaining thus an explicit \emph{cycle-by-cycle computation}. However, when low load levels are applied the crack growth rate is usually very low, leading to a long fatigue life characterized by a high number of cycles, e.~g. $N\ge10^5$ cycles. This regime takes the name of high cycle fatigue and is the most interesting for design purposes but also computationally time consuming. To reduce the computational time $\bs{\Delta} N$ can be set to an integer value greater than unity leading to a so-called \emph{cycle-jump approach} \cite{Cojocaru2006}. Although a rule for this simple approach cannot be stated, the value to assign to $\bs{\Delta} N$ usually decreases increasing the applied load range and, in general, it depends on the precision needed during the computation. More sophisticated approaches based on adaptive definition of $\bs{\Delta} N$ can be however found in the literature \cite{Cojocaru2006,Oskay2004,Bhattacharyya2019}.

\section{Numerical examples} \label{sct:num_ex}

\begin{figure}[!h]
	\begin{adjustwidth}{0cm}{0cm}
	\centering
		\includegraphics{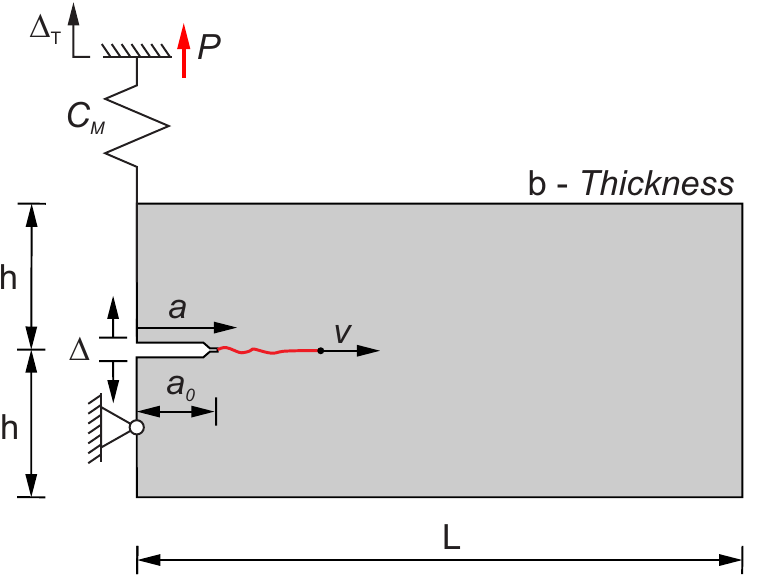}
	\end{adjustwidth}
		\caption{Geometry of the double cantilever beam test.}
		\label{fig:DCB}
	\end{figure}

	To test the capabilities of the proposed approach let us consider the double cantilever beam (DCB) specimen sketched in Fig.~\ref{fig:DCB} having dimensions $L\times 2h\times b$ and initial crack length $a_0$. The test is driven imposing the displacement $\Delta_{T,\,t}$ of a loading device connected to the specimen and with known compliance $C_M$. We assume that the arms of the DCB can be considered subjected to pure bending.
	
 The following dimensionless quantities are used
	\eqn{eq:dimensionless}{\begin{array}{cccc} \displaystyle\bar L = \frac{L}{L}=1\,, & \displaystyle\bar h = \frac{h}{L}\,, & \displaystyle\bar a = \frac{a}{L}\,, &\displaystyle\bar b = \frac{b}{L}\,, \\[20pt]
	\displaystyle\bar C_M = {C_M\gamma}\,,  & \displaystyle\bar Y = \frac{YL}{\gamma}\,, & \displaystyle\bar \Delta_T = \frac{\Delta_T}{L}\,, & \displaystyle\bar v= \frac{vT}{L} \,, \\ [20pt]%
						\displaystyle\bar G= \frac{G}{\gamma}\,, & \displaystyle\bar E= \frac{E}{\gamma L^2}\,, & \displaystyle\bar G_R = \frac{G_R}{\gamma}\,, & \displaystyle\bar t= \frac{t}{T}\,.					
							\end{array}}

\noindent where the Griffith critical energy release rate $\gamma$ and the time frame $T$ should be intended as reference values. For the geometry reported in Fig.~\ref{fig:DCB}, the dimensionless energy release rate is

	\eqn{eq:ERR_DCB_nondim}{\bar G(\bar \Delta_{T,\,\bar t},\,\bar a) = 12 \bar a^2 \bar Y \bar h^3\left[\frac{\bar \Delta_{T,\,\bar t}}{8\bar a^3+\bar C_M \bar Y \bar b \bar h^3}\right]^2\,,}

\noindent for a displacement-driven process, and 

	\eqn{eq:ERR_DCB_nondim_p}{\bar G(\bar P_{\bar t},\,\bar a) = \frac{12 \bar a^2}{\bar Y\bar b^2 \bar h^3}\bar P^2\,.}

\noindent  for a load-driven process \cite{Hutchinson1979}. Also, the dimensionless compliance $\bar C(\bar a)$ and applied load $\bar P(\bar \Delta_T(\bar t),\bar a)$ are

\eqn{eq:dimless_param}{ \bar C (\bar a)=\frac{8\bar a^3}{\bar Y \bar b \bar h^3}\,, \quad\quad \bar P_{\bar t}= \frac{\bar \Delta_{T,\,\bar t}}{\bar C(\bar a)+\bar C_M}\,.}

\noindent The dimensionless form of the loading ramp is taken as
		
		\eqn{eq:load_ramp}{\bar \Delta_{T,\,\bar t} = 10^{-3}\sqrt{\varepsilon \bar t}\,,}
		
\noindent where the dimensionless parameter $\varepsilon$ rules the loading rate \cite{Negri2010b}. This allows to define a rescaled time variable $\bar \tau$ as

	\eqn{eq:tau}{\bar \tau = \varepsilon \bar t\,.}

\noindent Note that the rate-independent case can be heuristically obtained for $\varepsilon\to$0, namely for vanishing loading rates.

	Although the present approach allows for any loading ramp, the specific choice of (\ref{eq:load_ramp}) renders the energy release rate function (\ref{eq:ERR_DCB_nondim}) linearly dependent on the time variable, which facilitate the comparison of the results.

If not otherwise specified all the results presented in the following sections are obtained using the parameters summarized in Tab.~\ref{tab:param}. Note, that imposing the same time increment in terms of rescaled time variable $\tau=\varepsilon\,t$ allows a fair comparison of the results since it ensures that the solution is computed for the same energy release rate independently on the loading rate. 

The reference solution in terms of crack length $\bar a$ for the rate-dependent case is obtained numerically integrating (\ref{eq:v_ODE}) with the Matlab algorithm \emph{ode15s} using a logarithmically spaced time discretization composed of 1000 points and a tolerance of $3\cdot10^{-14}$. The crack tip velocity is computed from the crack size vs. time evolution by means of  central finite differences. When relevant, the rate-independent reference solution is obtained analytically by solving the related problem as detailed in \cite{Carrara2020}. The data-driven solutions are obtained using artificially generated data sets created by random sampling of different underlying analytical relationships. Concerning the latter, in the rate-dependent fracture mechanics literature it is customary to consider the following additive decomposition of the resistance energy release rate curve $G_R(a,v)$ \cite{Fineberg1999,Lefranc2014}

	\eqn{eq:add_decomp}{G_R(a,v)=G_R^{RD}(a,v) + G_R^{QS}(a)\,,}

\noindent where $G_R^{RD}(a,v)$ is the rate-dependent non-decreasing contribution that vanishes for vanishing crack tip velocities.

\begin{table}
\centering
\begin{tabular}{llll}
\toprule
 Young's modulus  & $Y$ & = &  $70$ GPa\\[4pt]
 Height& $h$ & = & $3$ mm   \\[4pt]
 Length & $ L $& = &  $30$ mm  \\[4pt]
 Thickness & $ b $& = &  $1$ mm  \\[4pt]
 Initial crack length & $ a_0 $& = &  $3$ mm  \\[4pt]
Displacement ramp & $ \Delta_{T,\,t} $& = &  $3\cdot10^{-2}\sqrt{\varepsilon t}\,$ mm   \\[4pt]
Rescaled time increment & $ \bs{\Delta} \tau $& = &  $0.1$ sec   \\[4pt]
 Machine compliance &  $C_M$& = & $2\cdot 10^{-3}$ mm/N \\ [4pt]
Griffith fracture toughness &  \multirow{2}{*}{$\gamma$}& \multirow{2}{*}{=} & \multirow{2}{*}{$0.06$ N/mm} \\
\hspace{15mm}(Reference)   \\
Reference time frame &  {$T$}& {=} & {$1$ sec} \\
\toprule
\end{tabular}
\caption{Parameters used for the computations.}
\label{tab:param}
\end{table}

\FloatBarrier

	\subsection{Implicit Griffith-like model} \label{sct:Griff_ex}
	Discussed in this section are the results obtained using Algorithm~\ref{algo:DD_Griff} (Appendix~\ref{app:griffith}) and the following model featuring an implicit Griffith-like rate-independent limit
	
	\eqn{eq:Griff_model}{\bar G_R(\bar v)=1+2\bar v^2\,,}
	
\noindent where $\bar G_R^{QS}(\bar a)=G_c=1$ and $\bar G_R^{RD}(\bar v)=2\bar v^2$. The data sets used for the data-driven solution are obtained through a sampling of  (\ref{eq:Griff_model}) with 100 points randomly distributed along the interval $\bar v=[0, 4]$. 
	
		\subsubsection{Noiseless database} \label{sct:res_griff_noNoise}
		
\begin{figure}[!h]
	\begin{adjustwidth}{-3cm}{-3cm}
	\centering
		\includegraphics{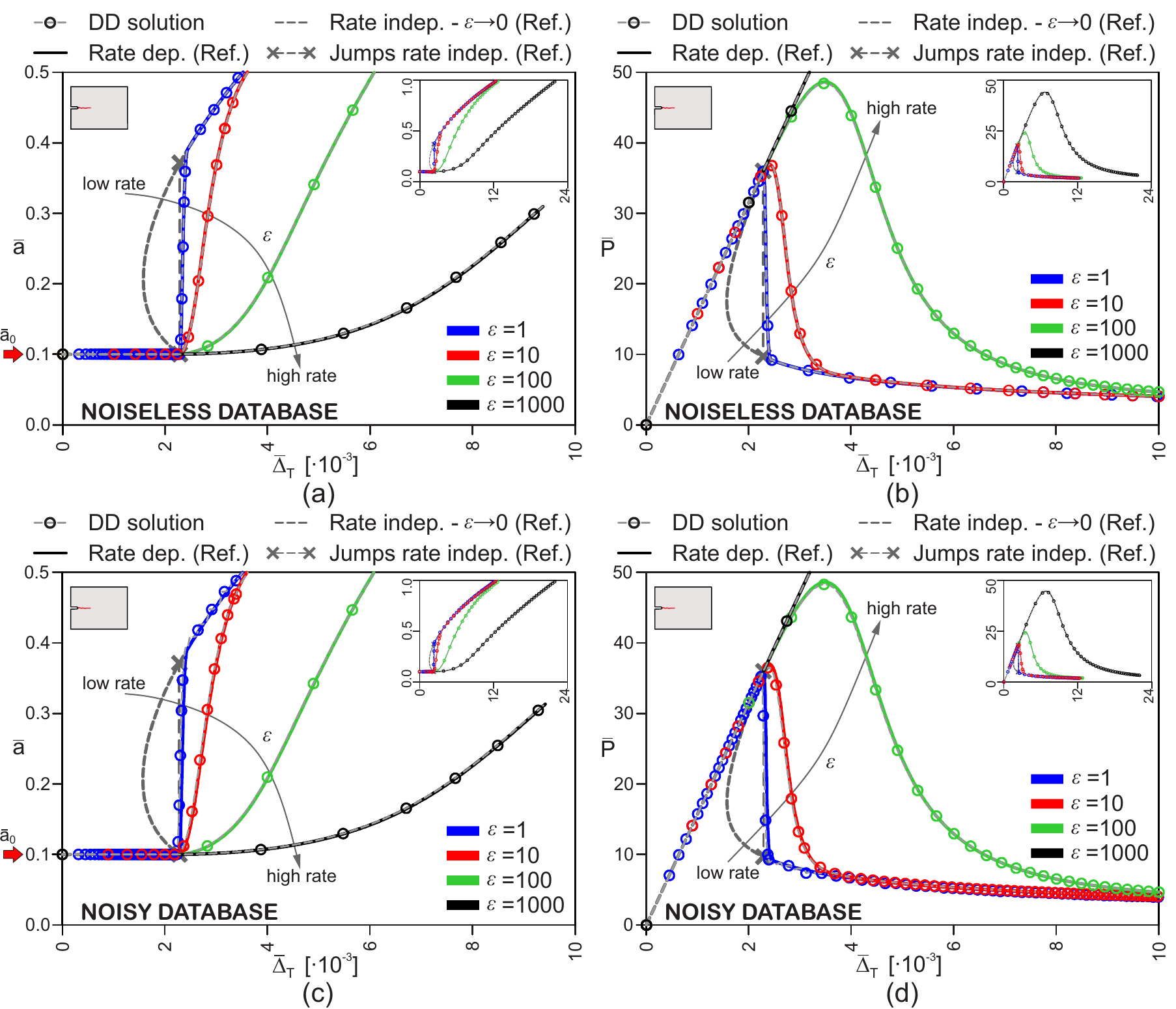}
	\end{adjustwidth}
		\caption{Comparison between reference and data-driven prediction for the implicit Griffith-like model for varying loading rate: results for the noiseless database in terms of crack length vs. displacement curves (a) and load vs. displacement curves (b); results for the noisy database in terms of crack length vs. displacement curves (c) and load vs. displacement curves (d).}
		\label{fig:Griff_comp}
	\end{figure}

		Figs.~\ref{fig:Griff_comp}a,b show the comparison between the reference and data-driven (DD in the figures) solution in terms of crack size and load evolution obtained with a noiseless database for different loading rate parameters $\varepsilon$=1, 10, 100 and 1000. 
		
\begin{figure}[!h]
	\begin{adjustwidth}{-3cm}{-3cm}
	\centering
		\includegraphics{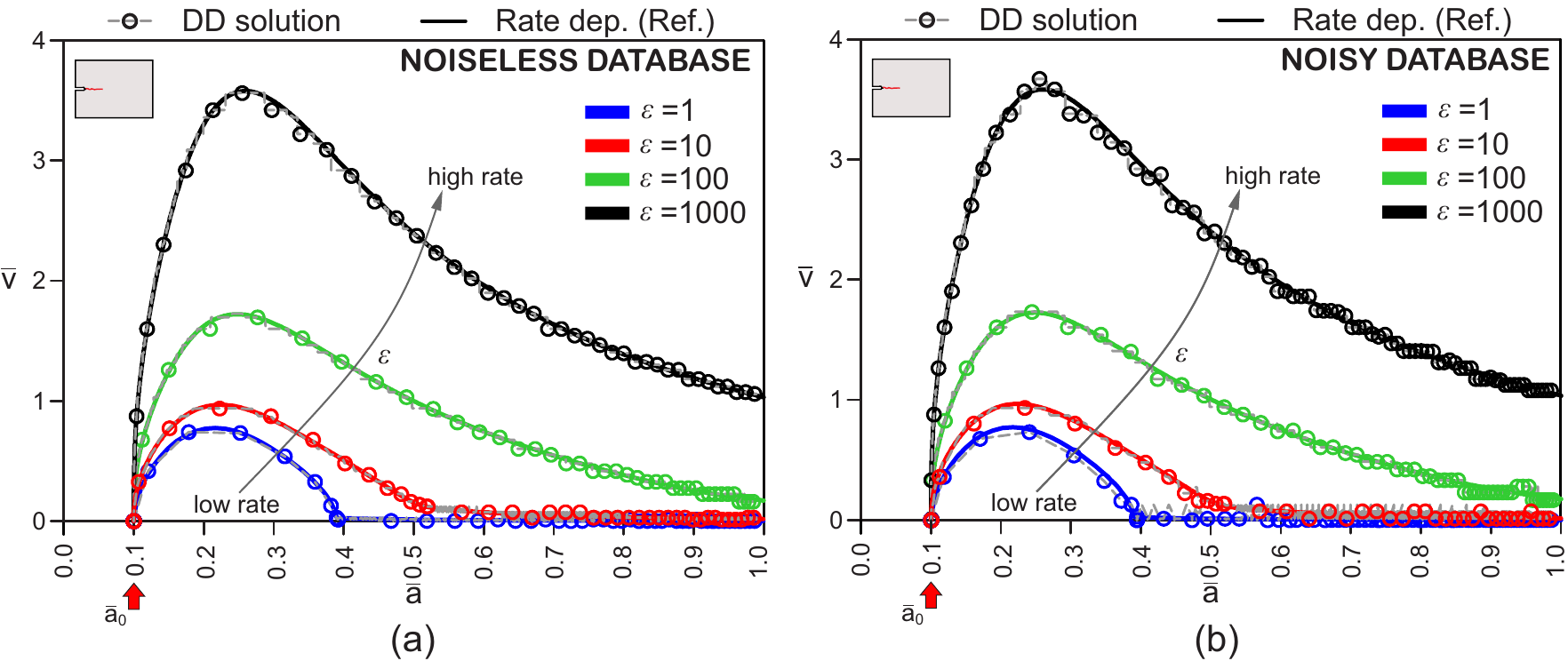}
	\end{adjustwidth}
		\caption{Comparison between reference and data-driven crack tip velocity profiles for the implicit Griffith-like model using a noiseless (a) and a noisy (b) database for varying loading rate.}
		\label{fig:Griff_comp_v}
	\end{figure}

		Considering the limited number of points in the data set, the agreement between reference and data-driven solution is excellent and, remarkably, the latter is able to reproduce the characteristic features of the rate-dependent fracture propagation. As expected, the evolutions of both crack size and load are drastically different for different loading rates \cite{Negri2010b}. In particular, the ultimate displacement increases with increasing loading rates (Fig.~\ref{fig:Griff_comp}a) and the same applies to the peak force (Fig.~\ref{fig:Griff_comp}b). Also, both reference and data-driven results converge toward the rate-independent result for sufficiently slow loading (Fig.~\ref{fig:Griff_comp}a,b). In the present case for $\varepsilon$=1 the rate-dependent solution is almost overlapped to the rate-independent limit. This is confirmed by Fig.~\ref{fig:Griff_comp_v}a where the crack tip velocity profiles along the crack path are shown. Here we can see that the crack tip velocity for $\varepsilon$=1 is close to zero for most of the test, while non negligible values are predicted only within the crack length range corresponding to the crack jump in the rate-independent case. Within such range and unlike in the rate-independent case, the rate-dependent framework and the proposed approach are able to continuously track the evolution of the system that is characterized by a rapidly evolving crack, a fast but smooth increase of the DCB compliance and, the drop of the applied load (Fig.~\ref{fig:Griff_comp}a,b). 
Afterwards, the system recovers the smooth and slow evolution of the rate-independent case.
	From Fig.~\ref{fig:Griff_comp_v}a we can also see how the crack tip velocity steadily increases increasing the loading rate parameter. Starting from $\varepsilon$=100 the rate-independent trend is no longer recovered even for very low values of the load (Fig.~\ref{fig:Griff_comp}a,b).
	
		\subsubsection{Noisy database} \label{sct:res_griff_noise}

		To determine the sensitivity of the proposed method to noise in the input data, a random white noise with amplitude $\pm$ 2.5\% in terms of difference between the observed and expected value of resistance energy release rate is applied to the material data set used in sect.~\ref{sct:res_griff_noNoise}.
		
		The obtained results and the comparison with the reference rate-dependent and -independent solutions are illustrated in Figs.~\ref{fig:Griff_comp}c,d and \ref{fig:Griff_comp_v}b in terms of crack size, load evolution and crack tip velocity profiles for the loading rate parameters $\varepsilon$=1, 10, 100 and 1000. Although the mismatch between data-driven and reference results is not as low as for the noiseless case, the agreement is still remarkable within the whole loading rate range tested and the observations of sect.~\ref{sct:res_griff_noNoise} apply. This confirms that the closest-point-projection strategy is a noise-wise robust approach to the data-driven fracture mechanics problem. 
		
		Note that, although remaining satisfactory, the agreement between data-driven and reference results slightly degrades toward the final part of the test especially in case of low loading rates (Fig.~\ref{fig:Griff_comp_v}b). Within such region, the crack front velocity becomes almost constant and, since the data-driven solution must be selected among the discrete points of the data set, the solver alternates under- and over-estimation of the velocity depending on the distribution of the material points. Note however, that the data-driven solution fluctuates closely  around the reference values (Fig.~\ref{fig:Griff_comp_v}b).  This discrepancy is averaged out when the velocity changes rapidly (Fig.~\ref{fig:Griff_comp_v}b). A similar behavior is also present using a noiseless data set (Fig.~\ref{fig:Griff_comp_v}a), however it is less visible there because the noise does not add up to the effect as in Fig.~\ref{fig:Griff_comp_v}b.

\begin{figure}[!h]
	\begin{adjustwidth}{-3cm}{-3cm}
	\centering
		\includegraphics{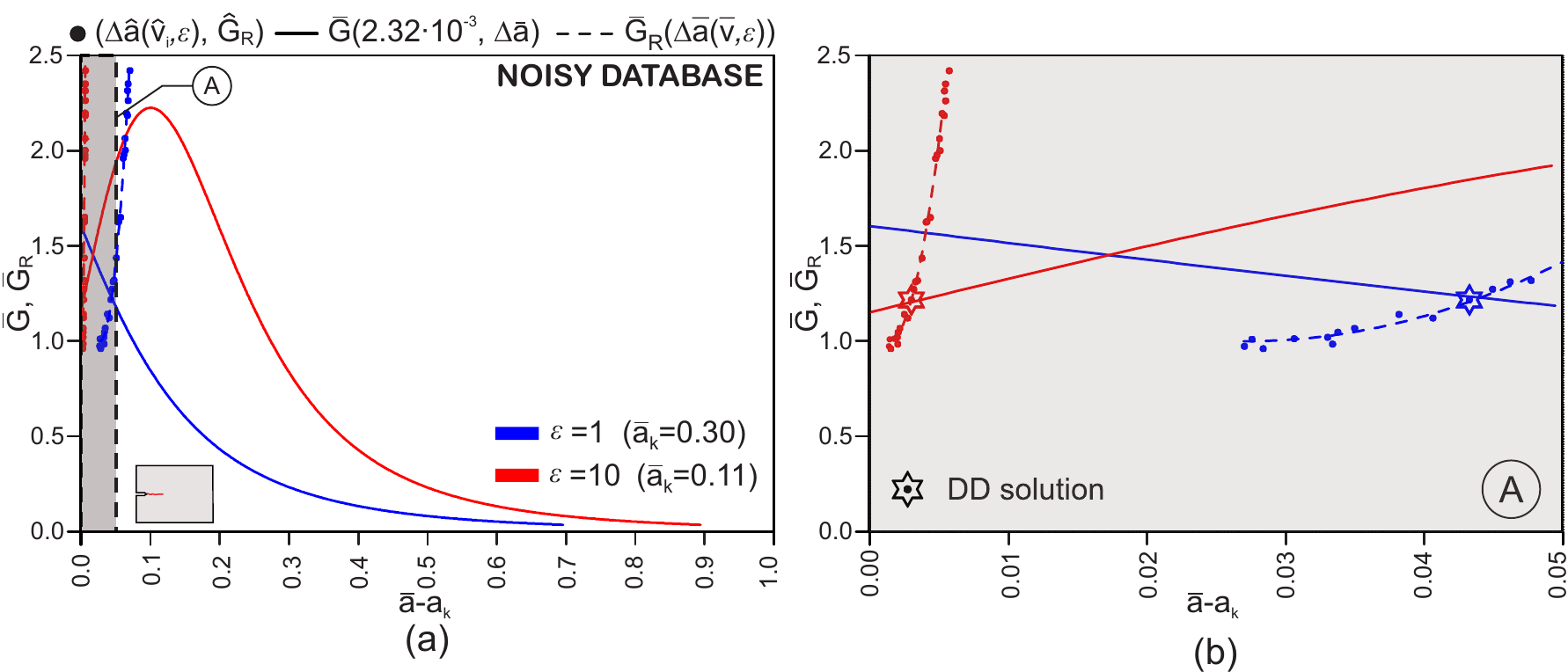}
	\end{adjustwidth}
		\caption{Data-driven search procedure at a device displacement $\bar \Delta_T$=2.32$\cdot$10$^{-3}$: (a) comparison between the energy release rate function and the adopted data set, (b) detail of the area $\textsf{A}$.}
		\label{fig:DD_search_noise}
	\end{figure}

	Fig.~\ref{fig:DD_search_noise} shows the data-driven search procedure for a device displacement $\bar \Delta_T$=2.32$\cdot$10$^{-3}$ for the loading rate $\varepsilon=1,\,10$. In particular, Figs.~\ref{fig:DD_search_noise}b illustrate how the current solution is selected as the point in the data set closest to the energy release rate function. The data are presented in the $(\bar G-\bs{\Delta} \bar a)$ plane and, although the resistance database is always the same, the set of points used for the closest-point-projection procedure changes depending on the loading rate parameter $\varepsilon$. This happens because the computations are performed keeping constant the rescaled time increment $\bs{\Delta} \bar \tau$ and, thus the crack increments $\bs{\Delta} \hat a$ that can be reached at a certain time step change as well depending on $\varepsilon$ and $\bar v_k$ following
	
	\eqn{eq:Da_eps}{\bs{\Delta} \hat a(\hat v_i,\,\varepsilon) = \frac{\bar v_k + \hat v_i}{2}\frac{\bs{\Delta} \bar \tau}{\varepsilon}\,.}
	
	\noindent In particular, we observe that decreasing the load rate parameter, the crack extensions that can be attained increase and that, following (\ref{eq:Da_eps}), for $\varepsilon\to$0 a horizontal distribution of the points of the data set would be reached, corresponding to a rate-independent Griffith resistance curve. In this case, the proposed approach reduces to the local minimization strategy based on closest-point-projection in \cite{Carrara2020}. Also, following (\ref{eq:Da_eps}), the minimum crack extension that can be reached in general does not vanish for vanishing crack tip velocities if the initial velocity $\bar v_k\neq 0$. 
		
	\FloatBarrier

		\subsubsection{Non-proportional loading}	
	
		Aim of this section is to demonstrate the capability of the proposed approach to correctly reproduce the reference results also in case of non-proportional  loading. The noisy material database adopted is the same used in sect.~\ref{sct:res_griff_noise}.
		
\begin{figure}[!h]
	\begin{adjustwidth}{-3cm}{-3cm}
	\centering
		\includegraphics{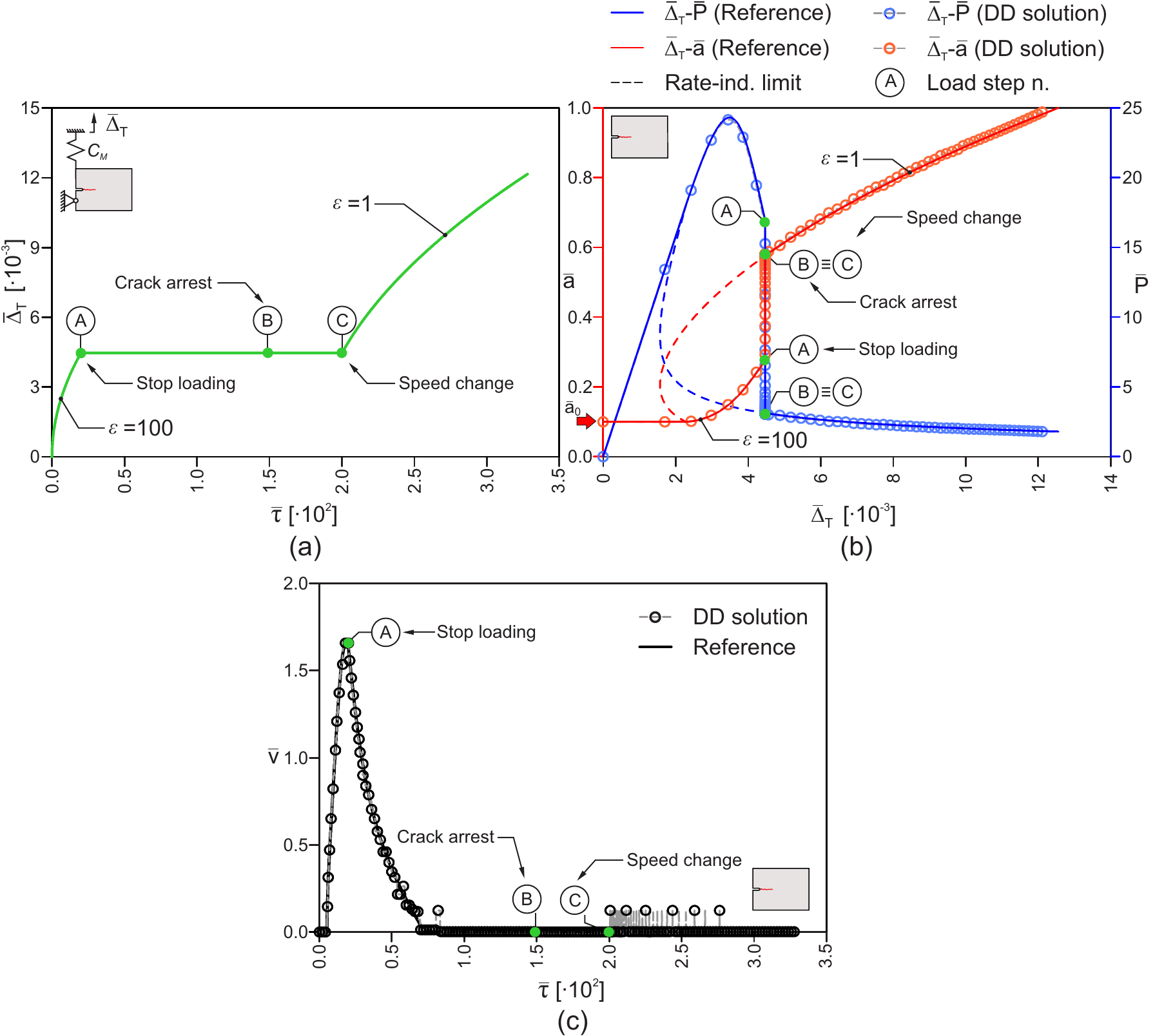}
	\end{adjustwidth}
		\caption{Comparison between reference and data-driven results for non-proportional loading with change of loading rate: (a) loading procedure, (b)  crack length vs. displacement and load vs. displacement curves and (c) velocity profile.}
		\label{fig:nonprop_1}
	\end{figure}

		The first test involves the loading ramp depicted in Fig.~\ref{fig:nonprop_1}a characterized by an initial fast loading with $\varepsilon$=100 until reaching a device displacement $\bar \Delta_T$=5$\cdot$10$^{-3}$ followed by a constant displacement branch up to $\bar \tau$=200 (i.~e., $\bar t$=2)  and then a slow loading with $\varepsilon$=1 until failure. Figs.~\ref{fig:nonprop_1}b,c show again an excellent agreement between reference and data-driven results. During the initial phase of the loading the system evolution significantly deviates from the rate-independent limit, then, upon stopping the loading increment (point \textsf{A}), the crack keeps evolving until finding its quasi-static equilibrium state at $\bar \tau\simeq$150 (point \textsf{B}, Fig.~\ref{fig:nonprop_1}b). At this point the crack arrests its evolution and the crack tip velocity remains zero until resuming the load increments (point \textsf{C}, Fig.~\ref{fig:nonprop_1}c). Right after this phase the crack tip velocity oscillates because of the discrete nature of the material data set along with the effect of the noise, that lead to the alternation of over- and under-estimations of the crack tip velocity.
	
\begin{figure}[!h]
	\begin{adjustwidth}{-3cm}{-3cm}
	\centering
		\includegraphics{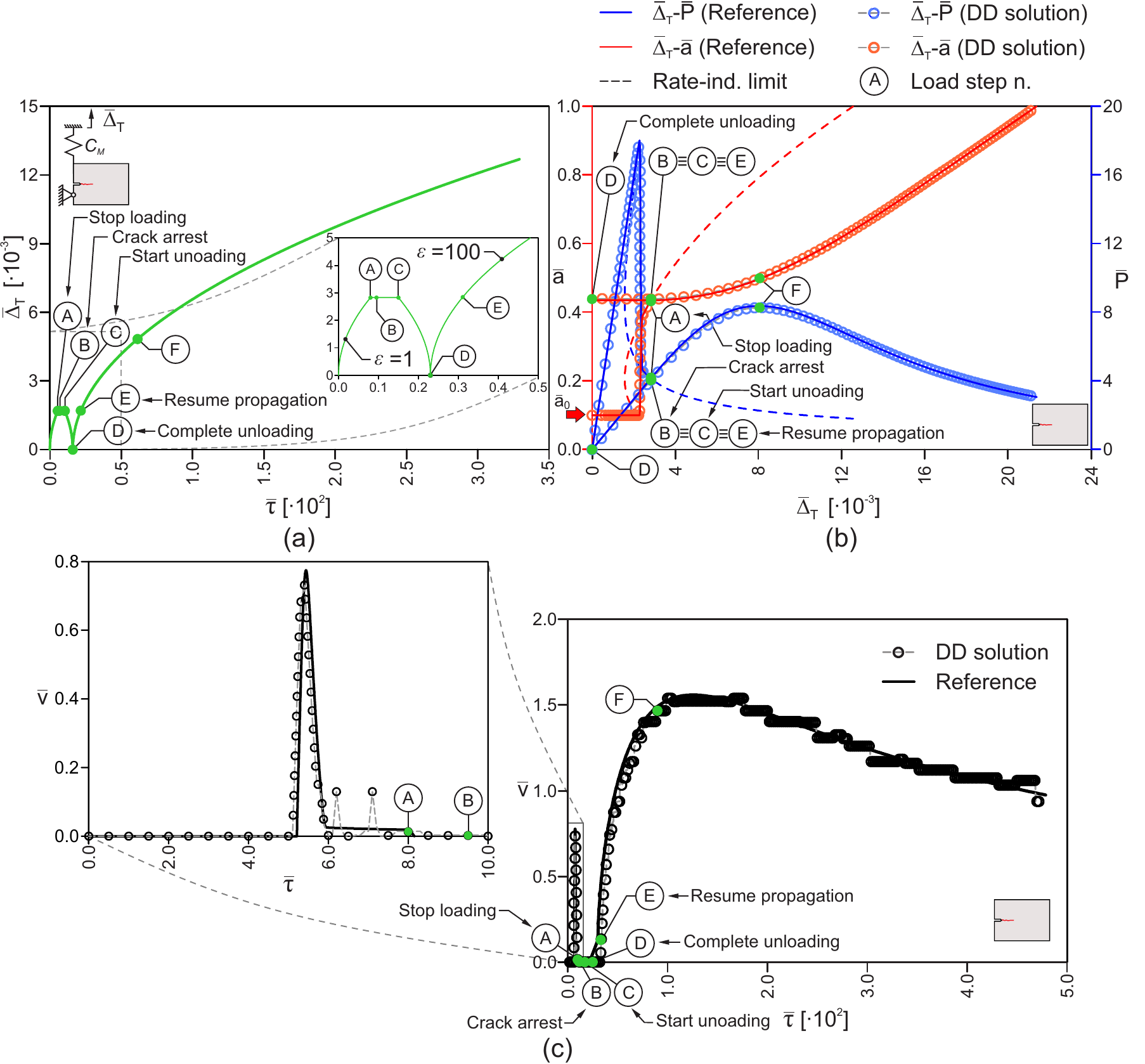}
	\end{adjustwidth}
		\caption{Comparison between reference and data-driven results for non-proportional loading  with change of loading rate and complete unloading: (a) loading procedure, (b)  crack length vs. displacement and load vs. displacement curves and (c) velocity profile.}
		\label{fig:nonprop_2}
	\end{figure}

	The second test involves a slow initial loading with $\varepsilon$=1 (until point \textsf{A}, Fig.~\ref{fig:nonprop_2}a), followed by a constant displacement branch (segment \textsf{A}-\textsf{C}, Fig.~\ref{fig:nonprop_2}a), a complete unloading (segment \textsf{C}-\textsf{D}, Fig.~\ref{fig:nonprop_2}a) and then a fast loading with $\varepsilon$=100 until failure. The comparison between data-driven and reference results is again excellent (Figs.~\ref{fig:nonprop_2}b,c). In particular, the solution initially follows closely the rate-independent results and this is confirmed also by the relatively limited time needed by the crack to arrest its evolution (point \textsf{C} in Figs.~\ref{fig:nonprop_2}b,c). Upon unloading and reloading, the system behaves as expected linear-elastically (Fig.~\ref{fig:nonprop_2}b), until resuming crack propagation after point \textsf{E}. This phase is followed by a second load sub-peak (point \textsf{F}) due to the higher loading rate effect that induces a relatively high crack tip velocity (Fig.~\ref{fig:nonprop_2}c). Since the value of $\hat G_{R,i}$ related to the minimum value of $\hat v_i$ that is used to initialize $\hat G_R^{QS}$ is affected by a negative noise, in Fig.~\ref{fig:nonprop_2}c we can observe that the initial crack propagation is slightly anticipated in the data-driven results.
		
Another common test featuring a non-proportional loading is used to characterize the rate-dependent behavior of, e.~g., soft materials \cite{Lefranc2014}. The test procedure involves the application of a certain displacement or load to a specimen without any initial pre-crack. Then, a sharp notch of length $a_0$ is created in the specimen while keeping the applied load or displacement constant and the system is let free to either relax until crack arrest or evolve until failure. 

\begin{figure}[!h]
	\begin{adjustwidth}{-3cm}{-3cm}
	\centering
		\includegraphics{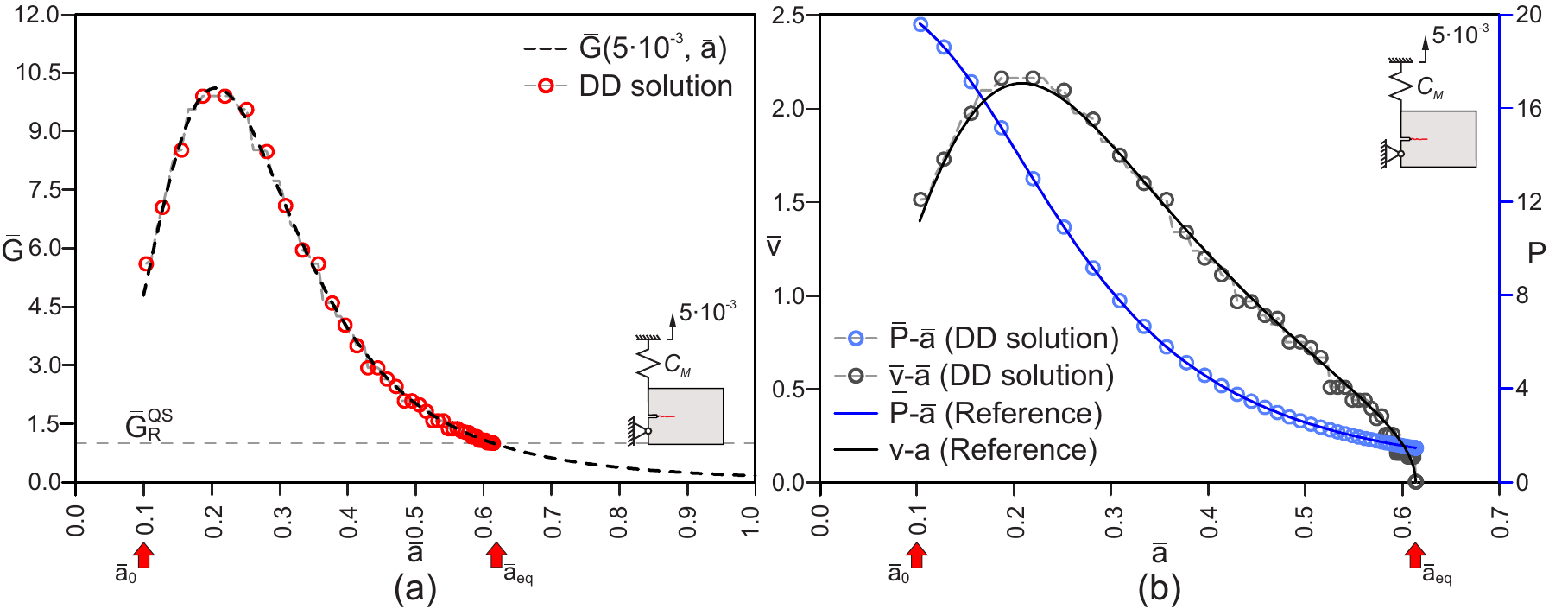}
	\end{adjustwidth}
		\caption{Constant displacement cracking test: (a) comparison between complete data-driven solution and reference energy release rate curve and (b) load and crack tip velocity profiles.}
		\label{fig:SM_u}
	\end{figure}

	Fig.~\ref{fig:SM_u} shows the results obtained for the constant displacement cracking test with an imposed displacement of $\bar \Delta_T$=5$\cdot$10$^{-3}$. We can see that the data-driven solution follows closely the energy release rate function obtained for the applied displacement until reaching an equilibrium crack size $\bar a_{eq}$ for which $\bar G(5 \cdot 10^{-3},\bar a_{eq}) \le \hat G_R^{QS}$ (Fig.~\ref{fig:SM_u}a). The agreement with the reference results is very good  in terms of both velocity and load (Fig.~\ref{fig:SM_u}b). Notably, the initial crack length $\bar a_0$ lies within the unstable branch of the energy release rate function and this leads to an initial increase of the crack tip velocity, that afterwards decreases until vanishing for $\bar a=\bar a_{eq}$  (Fig.~\ref{fig:SM_u}b). This evolution is also accompanied by a gradual relaxation of the load that reaches the equilibrium value following (\ref{eq:dimless_param}b) for $(\bar \Delta_T,\,\bar a)$=(5$\cdot$10$^{-3},\, \bar a_{eq})$. Note that before the crack arrest and for low crack tip velocities, the data-driven solution shows oscillations similar to what highlighted in sect.~\ref{sct:res_griff_noise} due to the discrete nature of the material data set (Fig.~\ref{fig:SM_u}b).
			
\begin{figure}[!h]
	\begin{adjustwidth}{-3cm}{-3cm}
	\centering
		\includegraphics{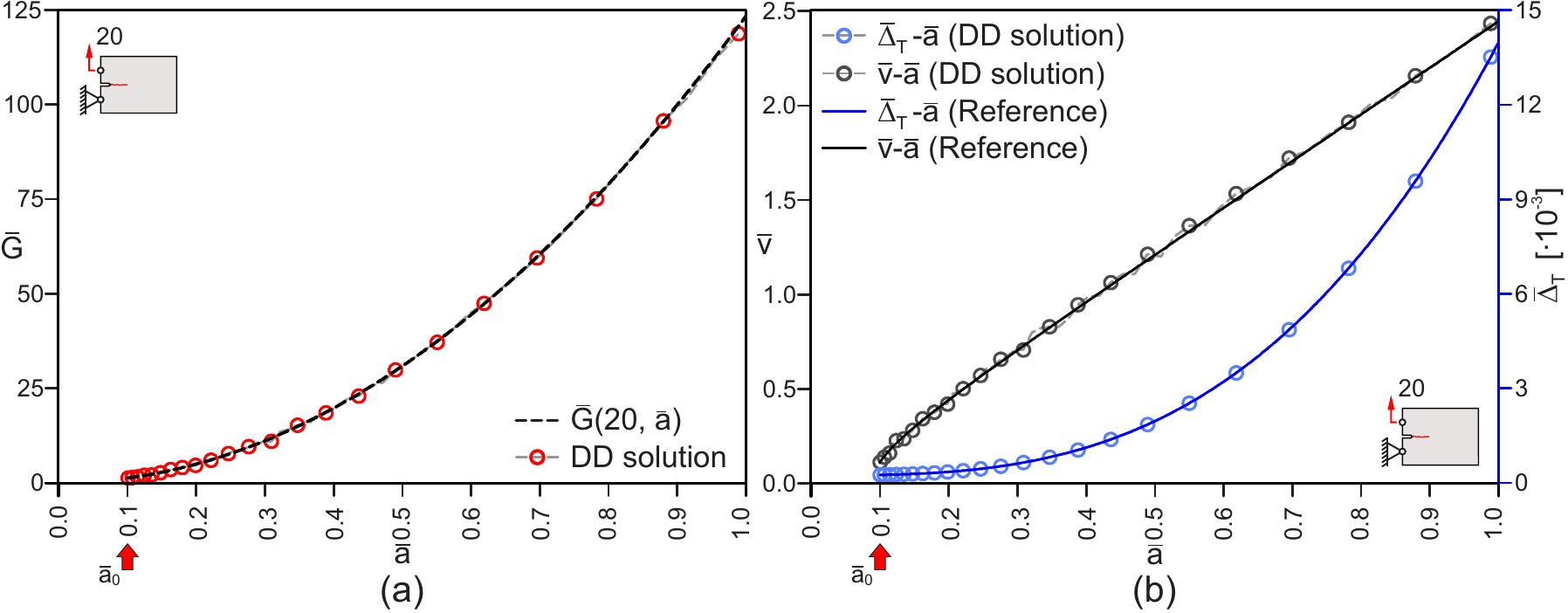}
	\end{adjustwidth}
		\caption{Constant load cracking test: (a) comparison between complete data-driven solution and reference energy release rate curve and (b) load and crack tip velocity profiles.}
		\label{fig:SM_P}
	\end{figure}

 		The results for a constant load cracking test with $\bar P$=20 are presented in Fig.~\ref{fig:SM_P}. In this case the energy release function is unstable for any value of $\bar a$ (Fig.~\ref{fig:SM_P}a) and this leads to a monotonic increase of the crack tip velocity until complete failure (Fig.~\ref{fig:SM_P}b). Note that, also in this case the data-driven approach proposed is able to correctly reproduce the reference results also in the final phases of the test where the crack tip velocity tends to diverge (Fig.~\ref{fig:SM_P}b).

	\FloatBarrier
	
	\subsection{Complete model}
	In this section we present the results for the complete model obtained following Algorithm~\ref{algo:DD_3D} (Appendix~\ref{app:complete}) and we compare the results with the rate-independent and the implicit Griffith-like model. All the examples involve a material data set composed of 300 and 50 points respectively for the rate-dependent and -independent material states, randomly sampled within the interval $(\bar v,\,\bar a)=([0,\,1.2],\,[0,\,1.1])$. Also, for the remainder of the paper a randomly assigned white noise with amplitude $\pm$ 2.5\% is applied similarly to sect.~\ref{sct:res_griff_noise}.

		\subsubsection{Comparison with the implicit Griffith-like model} \label{sct:Griff_3D}
		
\begin{figure}[!h]
	\begin{adjustwidth}{-3cm}{-3cm}
	\centering
		\includegraphics{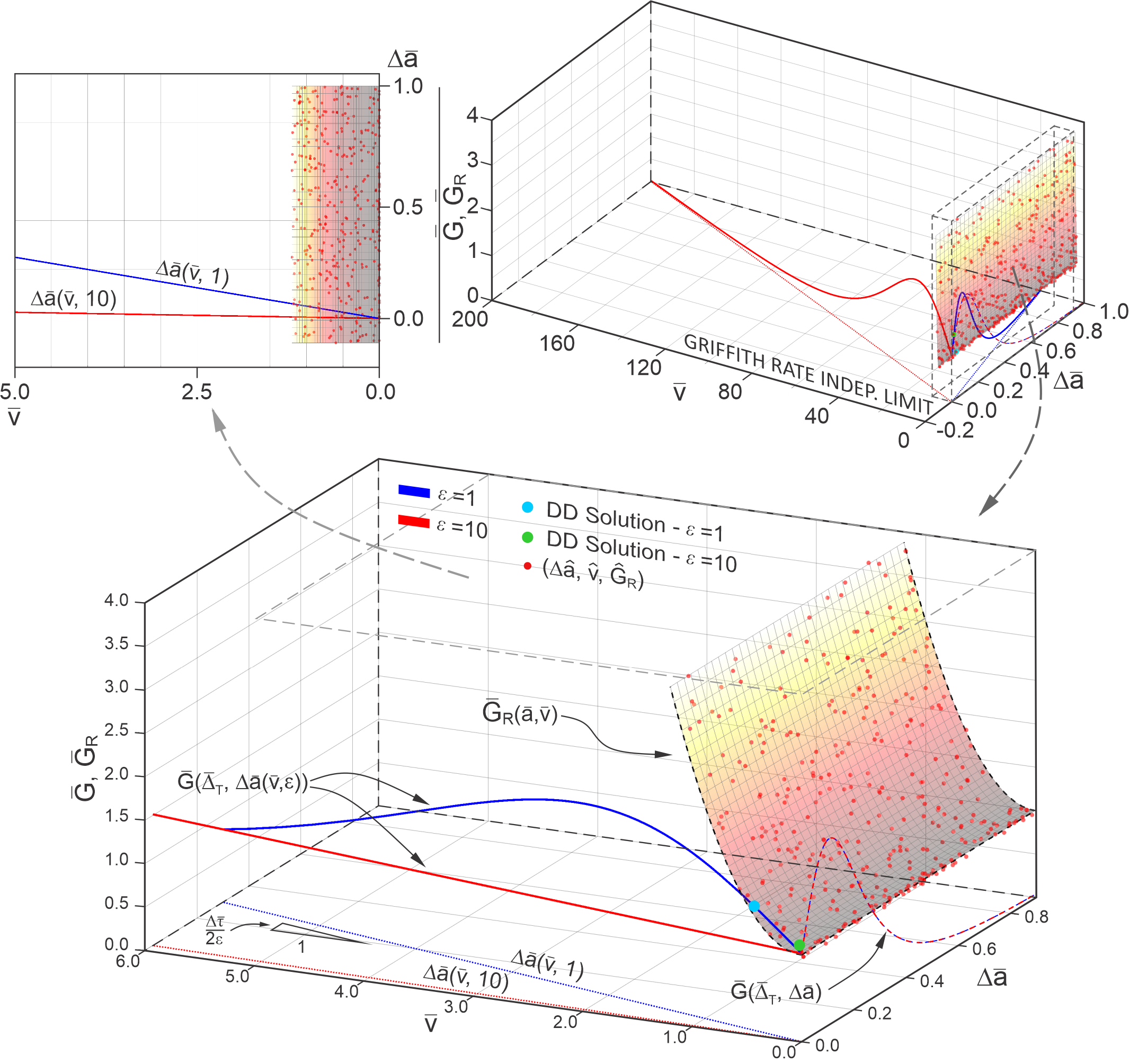}
	\end{adjustwidth}
		\caption{Data-driven search procedure for the complete model.}
		\label{fig:DD_search_3d}
	\end{figure}

		The first example aims at reproducing the results of the implicit Griffith-like model and, hence, adopts as underlying analytical model equation (\ref{eq:Griff_model}). Fig.~\ref{fig:DD_search_3d} shows the data-driven search procedure related to the first dissipative step and for loading rate parameter $\varepsilon$=1 and 10.

		Since the value of $\Delta \bar \tau$ is kept constant for all the computations, the plane over which the energy release rate function is projected depends on the value of $\varepsilon$ and its trace on the $\Delta \bar a-\bar v$ plane identifies, at the load step $k+1$, the line 
		\eqn{eq:line_3d}{\Delta \bar a_{k+1}(\bar v,\,\varepsilon)=\frac{\bar v_k+\bar v}{2}\frac{\Delta \bar \tau}{\varepsilon}\,.}
\noindent If $\varepsilon \to$ 0, (\ref{eq:line_3d}) coincides with the $\Delta \bar a$ axis, favoring the rate-independent solutions over the rate-dependent ones. Conversely, increasing $\varepsilon$ the lines become closer to the $\bar v$ axis, penalizing thus the rate-independent solutions (Fig.~\ref{fig:DD_search_3d}).

\begin{figure}[!h]
	\begin{adjustwidth}{-3cm}{-3cm}
	\centering
		\includegraphics{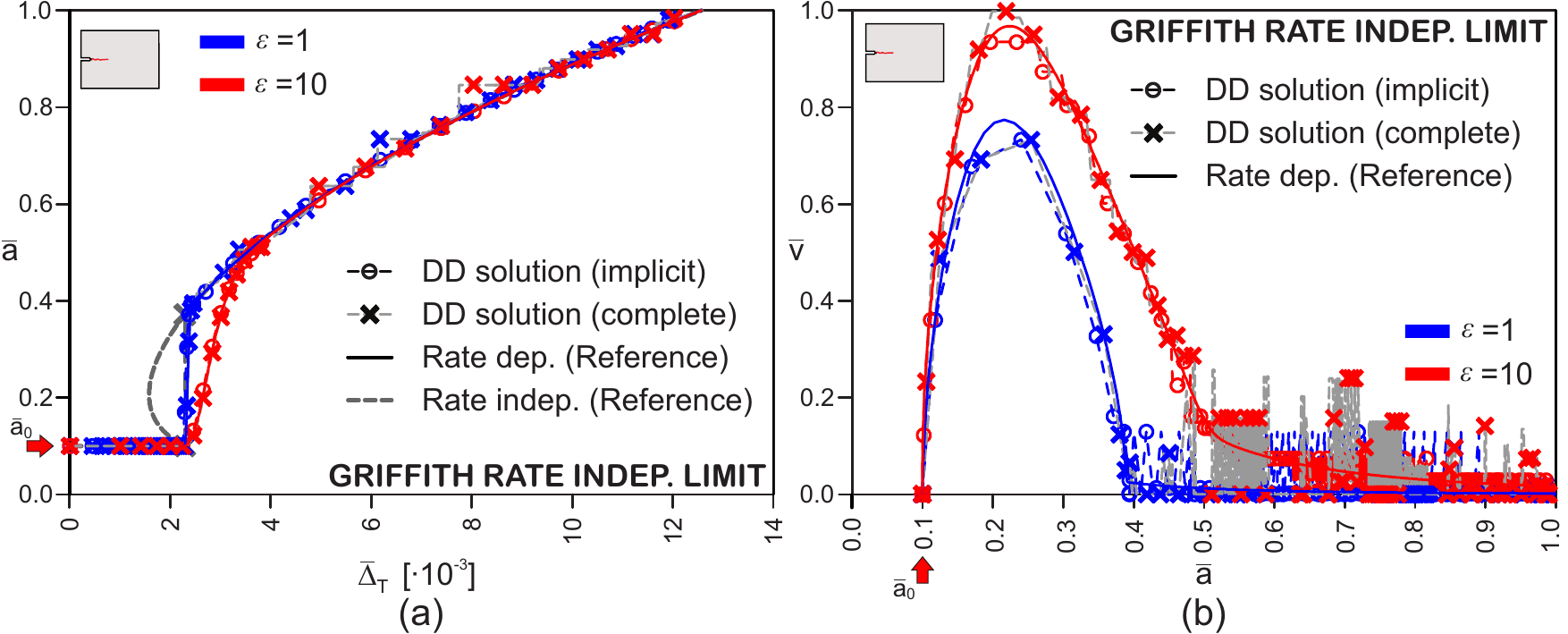}
	\end{adjustwidth}
		\caption{Comparison between reference results, implicit Griffith-like rate-independent limit model and complete model with a Griffith-like rate-independent limit: (a) crack size vs. displacement curve and (b) crack tip velocity profile.}
		\label{fig:3D_Griff}
	\end{figure}

		Fig.~\ref{fig:3D_Griff} presents the comparison between the reference solution and the data-driven results coming either from Algorithm~\ref{algo:DD_Griff} or \ref{algo:DD_3D} (Appendix~\ref{app:algorithms}). The agreement is excellent, considering that for the complete model data set there are wide unsampled areas (Fig.~\ref{fig:DD_search_3d}). The rate-independent solution is retrieved for sufficiently slow loading procedures (Fig.~\ref{fig:3D_Griff}a) and the reference crack tip velocity profile is well reproduced (Fig.~\ref{fig:3D_Griff}b), although in the final part of the test some oscillations are observable. This is again due to the presence of noise as well as to the discrete nature of the material data, leading to the alternation of dissipative steps with crack tip velocity higher than the reference one and elastic steps. Although taking place around the reference value, the magnitude of the oscillations in this case is higher than in the implicit Griffith-like case (sect.~\ref{sct:res_griff_noise}, Fig.~\ref{fig:Griff_comp_v}) especially for $\varepsilon$=10. This happens since, in this case, the density of the material points is lower.  Note also that, since the material data set is composed of rate-independent and -dependent states, in this case points with vanishing crack tip velocities are also present in the later stage of the test. 
	
		\subsubsection{R-curve rate-independent limit}
		Aim of this section is to show that Algorithm~\ref{algo:DD_3D} (Appendix~\ref{app:complete}) allows to reproduce generic behaviors depending also on the crack size and characterized by an R-curve quasi-static rate-independent limit. Here the following reference model is adopted
		\eqn{eq:3D_Rcurve}{\bar G_R(\bar a, \bar v) = \underbrace{1+\frac{(\bar a-0.1)^2}{(\bar a-0.1)^2+0.2(\bar a-0.1)}}_{\bar G_R^{QS}(\bar a)}+\underbrace{\vphantom{\frac{(\bar a-0.1)^2}{(\bar a-0.1)^2+0.2(\bar a-0.1)}}2\bar v^2}_{\bar G_R^{RD}(\bar v)}\,,}
\noindent where the terms $\bar G_R^{QS}(\bar a)$ and $\bar G_R^{RD}(\bar v)$ follow from (\ref{eq:add_decomp}).

\begin{figure}[!h]
	\begin{adjustwidth}{-3cm}{-3cm}
	\centering
		\includegraphics{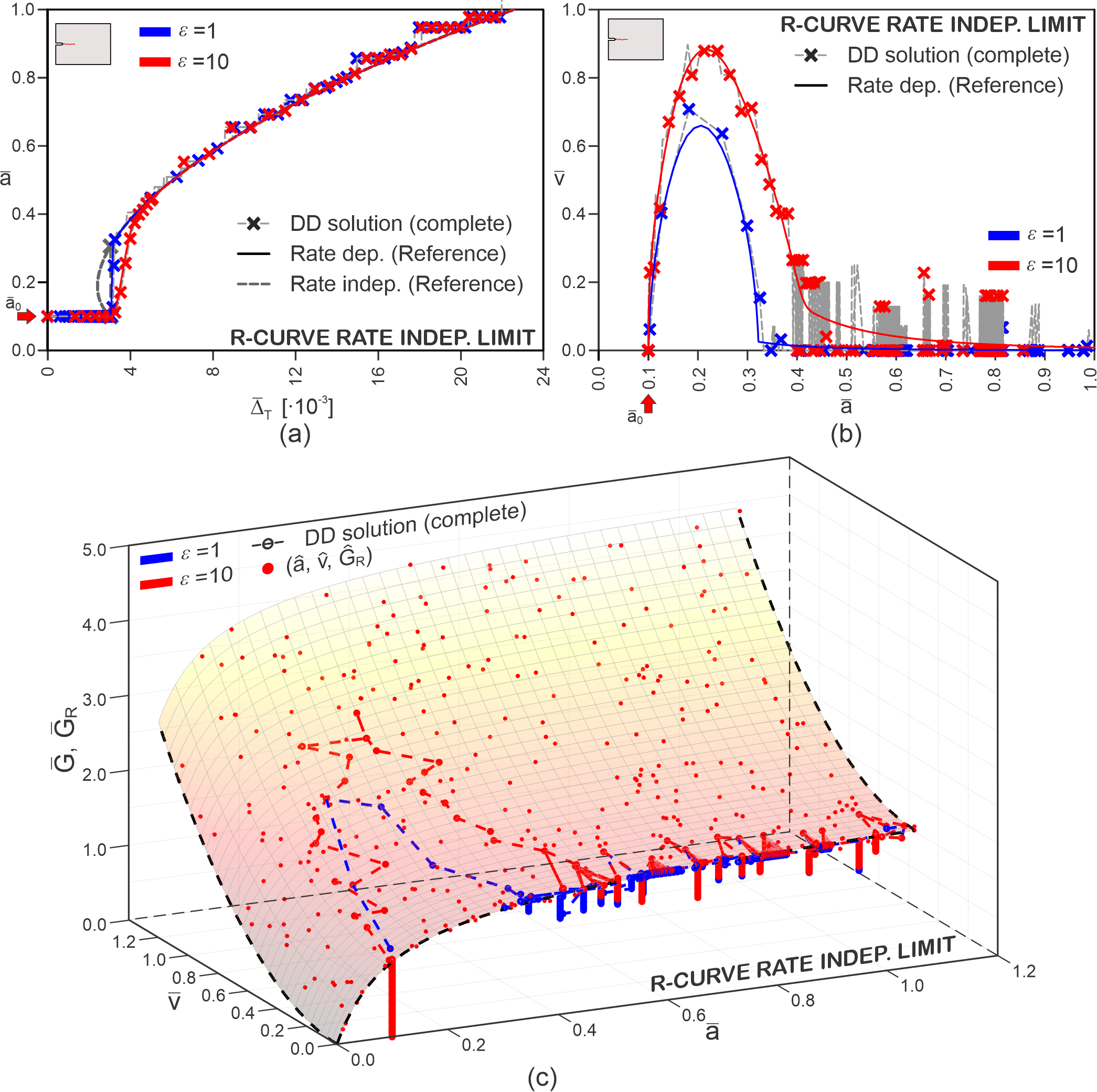}
	\end{adjustwidth}
		\caption{Comparison between reference and data-driven results using the complete model and a R-curve rate-independent limit for different loading rates: (a) crack size vs. displacement curve, (b) crack tip velocity profile and (c) complete data-driven solution.}
		\label{fig:comp_Rcurve}
	\end{figure}

		Figs.~\ref{fig:comp_Rcurve}a,b presents the comparison between reference and data-driven results for $\varepsilon$=1 and 10. The agreement is remarkable and similar to what highlighted in sect.~\ref{sct:Griff_3D}. Also in this case oscillations in the crack tip velocity profile are visible in the final part of the test (Fig.~\ref{fig:comp_Rcurve}b) that are wider for $\varepsilon$=10. Fig.~\ref{fig:comp_Rcurve}c, which illustrates the complete data-driven solution evolutions, confirms that within the region affected by crack tip velocity oscillation is characterized by the alternation of dissipative and linear elastic states. This means that, lacking a better solution, the data-driven procedure identifies as solution a crack tip velocity higher than the expected (reference) one. Hence, in the following steps the crack stops because the propagation condition is no longer satisfied although the applied displacement keeps increasing. The propagation then resumes when, after a few steps, the propagation condition is again satisfied.

		Fig.~\ref{fig:comp_Rcurve}c shows also that the proposed approach is able to let the fracture evolve either as a rate-dependent or -independent process. This is particularly evident for the case with $\varepsilon$=1 for a crack size $\bar a\ge$ 0.4 where the data-driven search algorithm mostly identifies material states that lie in the quasi-static rate-independent set, namely belonging to the plane $(\Delta \bar a\,-\,\bar G)$.
		
\begin{figure}[!h]
	\begin{adjustwidth}{-3cm}{-3cm}
	\centering
		\includegraphics{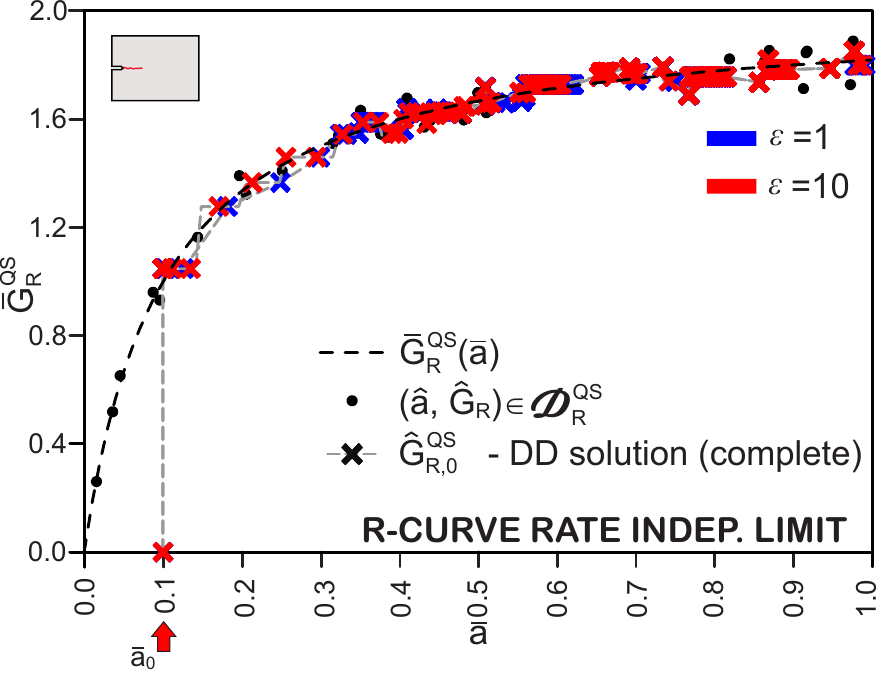}
	\end{adjustwidth}
		\caption{Comparison between reference and data-driven values of the rate-independent R-curve limit.}
		\label{fig:G0_Rcurve}
	\end{figure}

		Fig.~\ref{fig:G0_Rcurve} shows the profile of identified values of $\hat G_R^{QS}$ following (\ref{eq:avg_GR}) and, for comparison, the set of rate-independent states $\mathcal{D}_R^{QS}$ and the analytical $\bar G_R^{QS}(\bar a)$ curve of (\ref{eq:3D_Rcurve}). Here we can see that the proposed approach is able to closely follow the reference curve despite the limited amount of points and the noise. Note that, contrarily to the pure rate-independent case \cite{Carrara2020}, during crack propagation the reference value for the critical energy release rate is not always identified with a point in $\mathcal{D}_R^{QS}$ but it usually takes values of energy release rate between separate material states.

	\subsection{Regularization of the rate-independent problem} \label{sct:regularization}

	This section demonstrates how the algorithms described in sect.~\ref{sct:num_impl} can be conveniently adopted to regularize some pathological behaviors arising when adopting the rate-independent approach as shown in \cite{Carrara2020}. Two cases will be analyzed, namely the unstable tapered and bimaterial DCB tests. The rate-independent solutions of these setups are characterized by a non-convex free energy that includes multiple snap-back branches and the presence of multiple competing meta-stable states. Among these, the data-driven rate-independent solution selects the one closest to the energy release rate function, regardless of whether energetic barriers separate the initial and final states. 
	
	Since the introduction of realistic rate-dependent resistance models convexifies the free energy function, this approach is a good candidate to amend the aforementioned issues. Thus, the rate-dependent model can be seen as a viscous regularization of the non-convex minimization problem and, in this sense, goes in the direction of the vanishing viscosity approaches adopted, e.~g., by \cite{Alessi2018,Toader2009,Knees2008}. The assumptions on $G_R(a,\,v)$ that render the problem convex are inspired from the available experimental data, hence, upon substitution of the analytical constitutive laws with the discrete raw data, they are encoded into the data set.

		\subsubsection{Unstable tapered DCB}
		
\begin{figure}[!h]
	\begin{adjustwidth}{0cm}{0cm}
	\centering
		\includegraphics{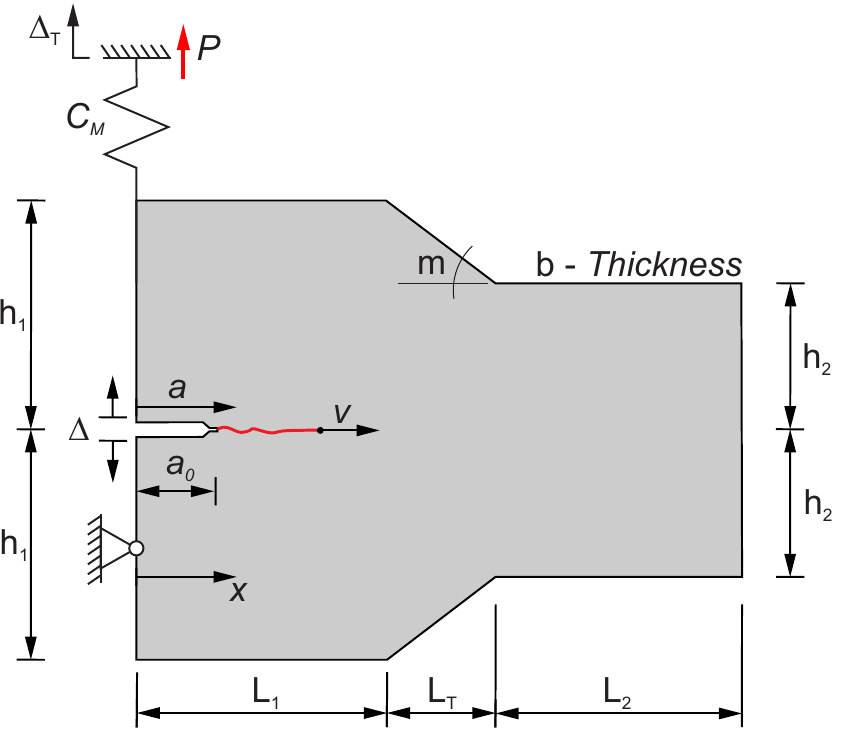}
	\end{adjustwidth}
		\caption{Scheme and geometry of the tapered double cantilever beam test.}
		\label{fig:taper_geom}
	\end{figure}

		The geometry of the setup adopted here is sketched in Fig.~\ref{fig:taper_geom}. The parameters adopted are the same of the unstable case labeled \textsf{case 3} in \cite{Carrara2020}, which are summarized in Tab.~\ref{tab:unst_taper}. The analytical compliance function is given in \cite{Carrara2020}. The same noisy material database used in sect.~\ref{sct:res_griff_noise} stemming from (\ref{eq:Griff_model}) is adopted. 
	
\begin{table}[!h]
\centering
\begin{tabular}{cccccc}
\toprule
$\bar h_1$ & $\bar h_2$ & $\bar L_1$ & $\bar L_T$ &  $\bar L_2$ &  $m$ ($^\circ$) \\[4pt]
\toprule
 0.10 & 0.04  & 0.45 & 0.10  & 0.45 & -3/5 (-30.96$^\circ$) \\
\toprule
\end{tabular}
\caption{Geometric parameters for the \textsf{case 3} tapered DCB in \cite{Carrara2020}.}
\label{tab:unst_taper}
\end{table}

\begin{figure}[!h]
	\begin{adjustwidth}{-3cm}{-3cm}
	\centering
		\includegraphics{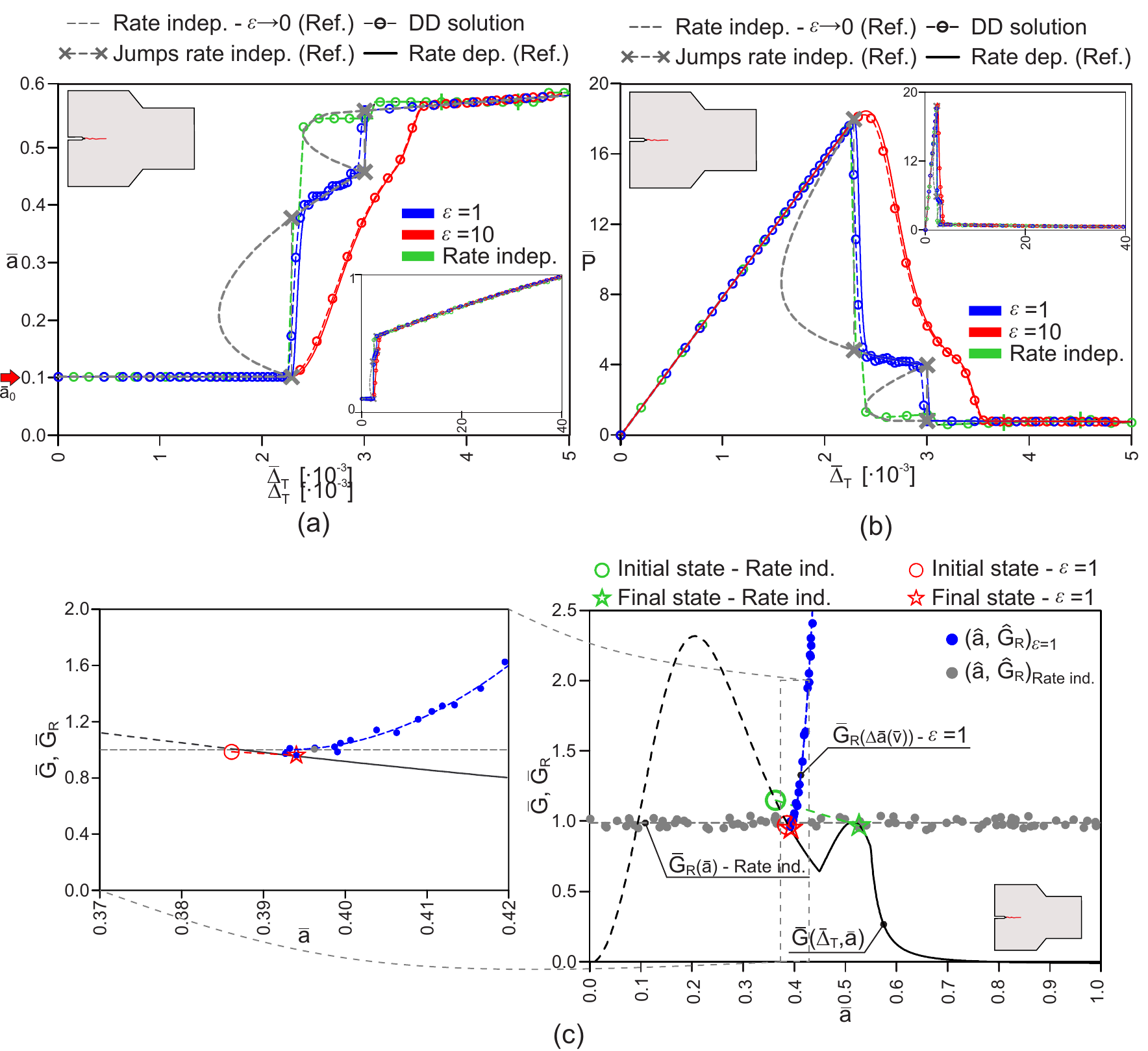}
	\end{adjustwidth}
		\caption{Comparison between data-driven results for the unstable tapered DCB specimen in case of rate-independent \cite{Carrara2020} and rate-dependent fracture propagation for different loading rates: (a) crack length vs. displacement curves, (b) load vs. displacement curves and (c) data-driven search procedure at the load step where the rate-independent model fails in detecting the local minimum.}
		\label{fig:res_taper}
	\end{figure}

	Figs.~\ref{fig:res_taper}a,b compares the results obtained for a loading rate parameter $\varepsilon$=1 and 10 with those obtained for the rate-independent case in \cite{Carrara2020} and the corresponding reference curves. The agreement between reference and data-driven results in the rate-dependent case is similar to what highlighted in sect.~\ref{sct:Griff_ex} and the same observations apply. Comparing the data-driven rate-dependent and -independent results it is possible to see that, unlike in the latter case, in the former the results for $\varepsilon$=1 closely approximate the rate-independent reference curves (Fig.~\ref{fig:res_taper}a,b). In particular, the proposed approach predicts, without the introduction of any \emph{ad-hoc} criterion, a steep but smooth crack (and hence load) evolution in correspondence of both rate-independent crack jumps, while before and after them it follows a gradual and stable crack evolution. As observed for the simple DCB test, the system evolution for $\varepsilon$=10 is smoother with a slightly higher load peak compared to the other case. 
	
	Fig.~\ref{fig:res_taper}c shows the comparison between closest point projection rate-dependent and -independent data-driven search procedure at $\bar \Delta_T $=2.4$\cdot$10$^{-3}$. As observable, the proposed approach amends the issues highlighted in \cite{Carrara2020} concerning the competition between different meta-stable states and it allows to identify a solution in close agreement with the reference one (Fig.~\ref{fig:res_taper}c). Of course, this is not guaranteed if an inappropriate very large time step is selected, so that the crack size increments become very large even for very low values of the crack tip velocity. In fact, the higher the time step, the more the material data set flattens, resembling, in the limit for $\Delta \bar t \to \infty$ (i.~e., for $\varepsilon\to 0$), the rate-independent one.

		\subsubsection{Unstable bimaterial DCB}
		
\begin{figure}[!h]
	\begin{adjustwidth}{0cm}{0cm}
	\centering
		\includegraphics{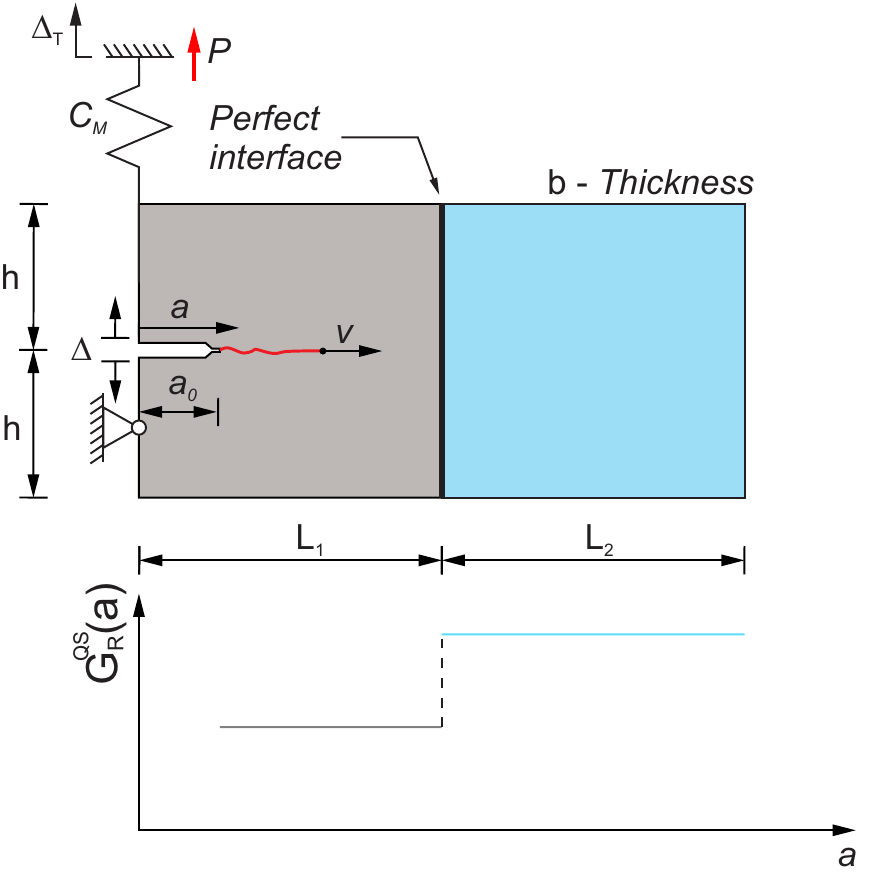}
	\end{adjustwidth}
		\caption{Scheme and geometry of the bimaterial double cantilever beam test.}
		\label{fig:bimat_geom}
	\end{figure}

		We consider now the setup illustrated in Fig.~\ref{fig:bimat_geom}, where the DCB specimen is composed of two different material connected by a perfect interface. As in \cite{Carrara2020}, the extensions of the two sections of the specimen are $\bar L_1=\bar L_2=0.5$, while the underlying unstable energy release rate resistance model is written following (\ref{eq:add_decomp}) with

		\eqn{eq:bimat_fct}{\bar G_R^{RD}(\bar v)=2\bar v^2\quad \text{and} \quad \bar G_R^{QS}(\bar a)=\begin{cases} 5 &\text{for } 0.0< \bar a \le 0.5\\
														 1 & \text{for } 0.5< \bar a \le 1.0\,.  \end{cases}}

\begin{figure}[!h]
	\begin{adjustwidth}{-3cm}{-3cm}
	\centering
		\includegraphics{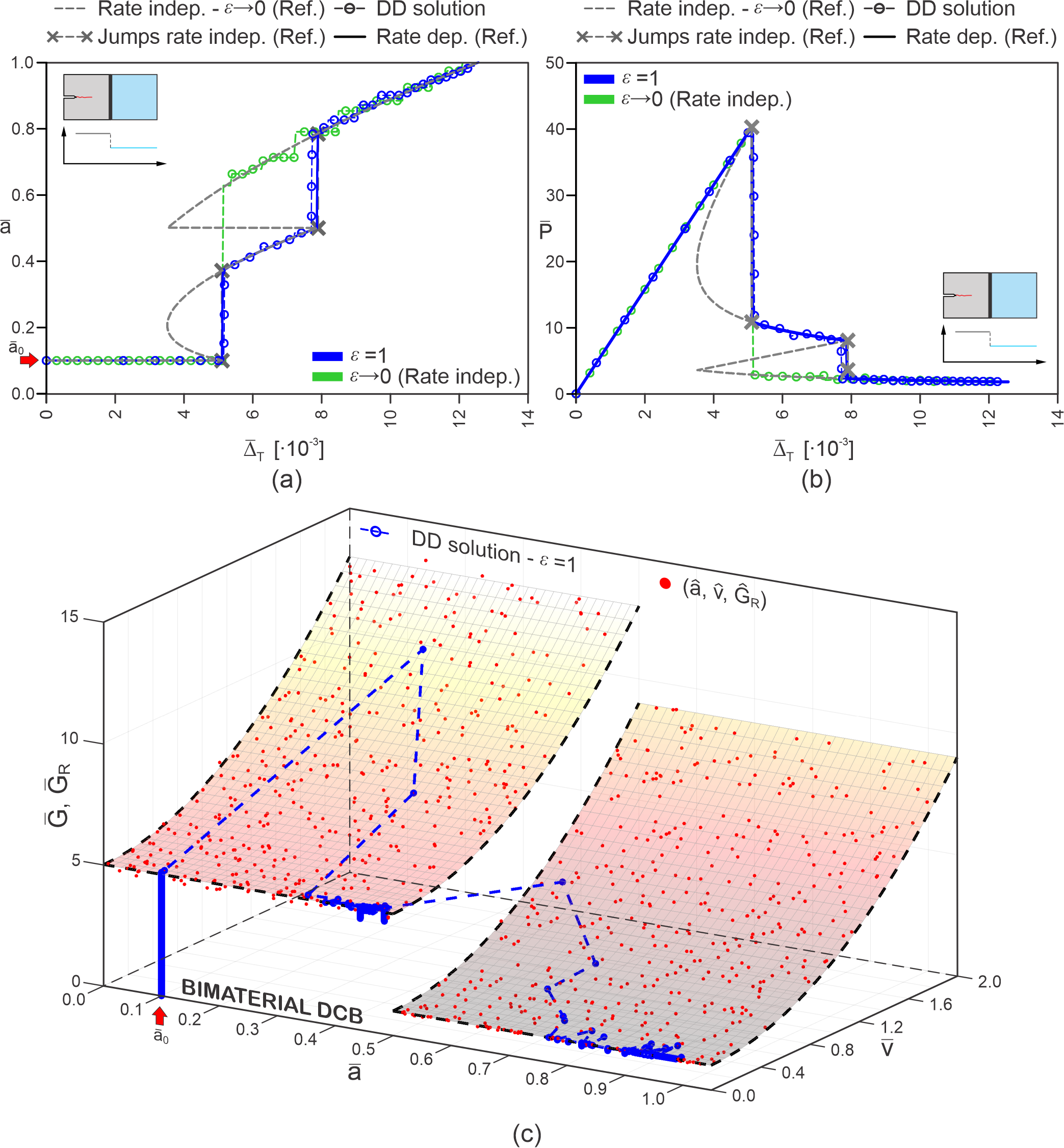}
	\end{adjustwidth}
		\caption{Comparison between data-driven results for the unstable bimaterial DCB specimen in case of rate-independent \cite{Carrara2020} and rate-dependent fracture propagation: (a) crack length vs. displacement curves, (b) load vs. displacement curves and (c) complete data-driven solution for the rate-dependent case.}
		\label{fig:bimat_result}
	\end{figure}

		Since in this case the rate-independent quasi-static resistance model depends on the crack extension $\bar a$ the complete model must be adopted. The obtained results for $\varepsilon$=1 are illustrated and compared to the reference and rate-independent results from \cite{Carrara2020} in Fig.~\ref{fig:bimat_result}. A noisy material database $\mathcal{D}_R$ is obtained by random sampling of (\ref{eq:bimat_fct}a) and (\ref{eq:bimat_fct}b) within the range $(\bar a,\, \bar v)=([0,\,1.1],\,[0,\,2])$ respectively with 700 and 100 points (Fig.~\ref{fig:bimat_result}c).
		
	Also in this case the proposed approach is able to overcome the issues related to the adoption of the rate-independent model (Figs.~\ref{fig:bimat_result}a,b). In particular, the first snap back branch is correctly reproduced along with the second one, which takes place when the crack meets the interface between the two materials. At this point, the crack tip velocity experiences a sudden increase and is thus forced to rapidly evolve inside the weaker portion of the specimen (Fig.~\ref{fig:bimat_result}c). Then the crack tip velocity gradually decreases reaching again values close to zero. On the contrary, the rate-independent solution involves a single long crack jump that starts as soon as the propagation conditions are met at $\bar a = \bar a_0$ and it immediately jumps inside the weaker material up to $\bar a\simeq$ 0.65 (Figs.~\ref{fig:bimat_result}a,b).

	\subsection{Fatigue}
	
	Within this section we explore the capability of the proposed approach to reproduce also the fatigue and sub-critical crack propagation behavior by adopting Algorithm~\ref{algo:DD_fat} (Appendix~\ref{app:fatigue}). As underlying fatigue constitutive law we adopt the \emph{NASGRO} law in the form \cite{Rabold2013}
	
\eqn{eq:nasgro}{\frac{da}{dN} =v_f= C_f\bs{\Delta} K^m \frac{\left( 1- \displaystyle\frac{K_T}{\bs{\Delta} K}\right)^p}{\left( 1- \displaystyle\frac{\bs{\Delta} K}{ K_c}\right)^q}= C_f(Y^\prime \bs{\Delta} G)^{m/2} \frac{\left( 1- \displaystyle\sqrt{\frac{G_T}{\bs{\Delta} G}}\right)^p}{\left( 1- \displaystyle\sqrt{\frac{\bs{\Delta} G}{ G_c}}\right)^q}\,,}

\noindent where the relation $\bs{\Delta} K = \sqrt{Y^\prime\,\bs{\Delta} G}$ is used and $C_f$, $m$, $p$ and $q$ are material parameters defining the shape of the law, while $K_c$ ($G_c=K_c^2/Y^\prime$) and $K_T$ ($G_T=K_T^2/Y^\prime$) are the critical and fatigue-threshold stress intensity factors (energy release rates) respectively. The material parameters adopted are summarized in Tab.~\ref{tab:NASGRO_param}. The dimensionless curve obtained using (\ref{eq:dimensionless}) and 

\begin{figure}[!h]
	\begin{adjustwidth}{-3cm}{-3cm}
	\centering
		\includegraphics{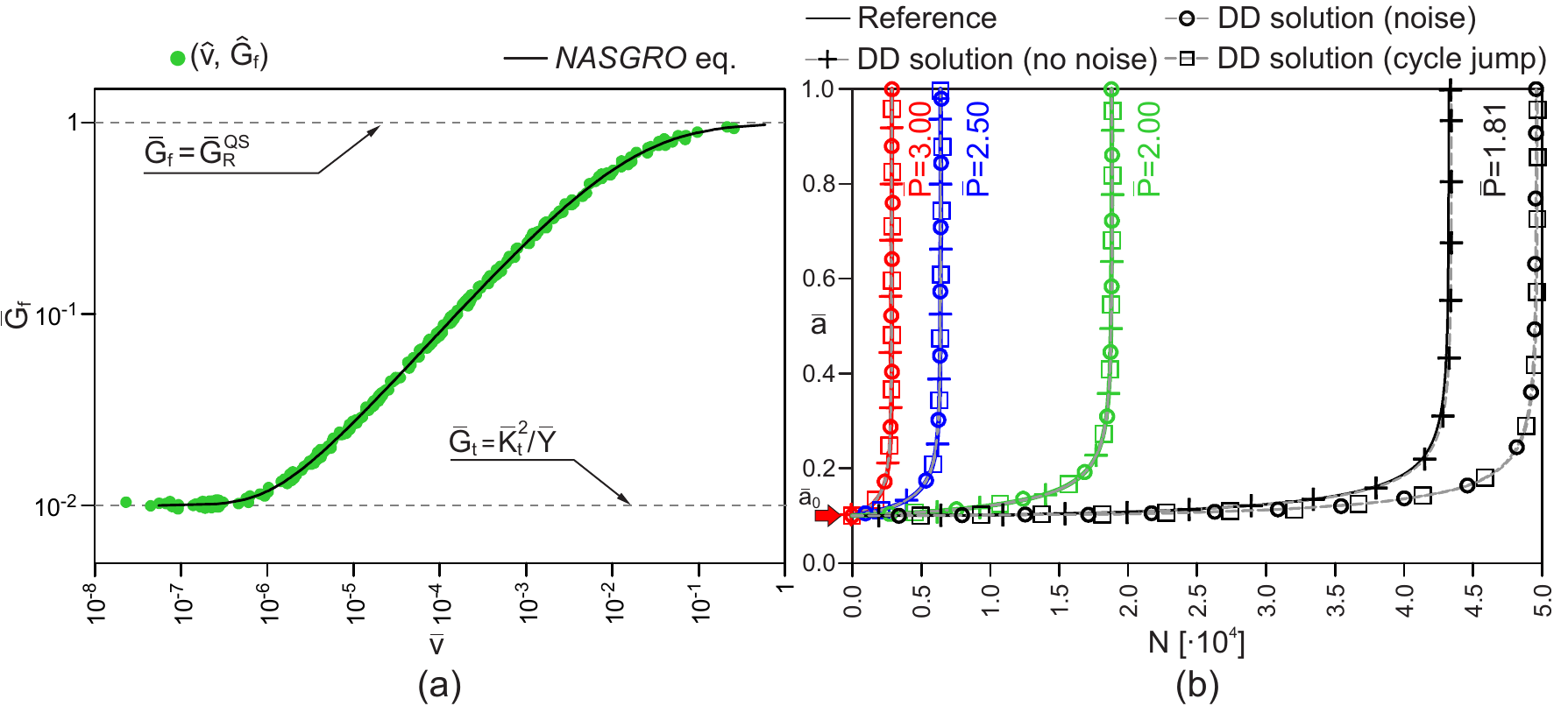}
	\end{adjustwidth}
		\caption{Sub-critical and fatigue crack growth: (a) noisy database adopted and (b) comparison between reference results and the data-driven predictions using a cycle-by-cycle simulation with a noiseless and noisy databases and a cycle-jump simulation with $\bs{\Delta} N$=10 and a noisy database. }
		\label{fig:comp_fat}
	\end{figure}

	\eqn{eq:dimensionless_fat}{\begin{array}{cccc} \displaystyle\bar K = \frac{K}{\sqrt{Y^\prime\gamma}}\,, & \displaystyle\bar C_f = \frac{C_f\gamma^m}{L^{(1+0.5m)}}\,.					
							\end{array}}
\noindent is shown in Fig.~\ref{fig:comp_fat}a along with the noisy material data set adopted, which is composed of 300 points randomly selected within the range $\bar G_{f} = [G_t/\gamma,\,G_c/\gamma]= [10^{-2},\,1]$. The same sampling of the energy release rate is used to obtain also a noiseless data set (not shown here).

\begin{table}[!h]
\centering
\begin{tabular}{cccc}
\toprule
 $C_f$  & &&10$^{-7}$ MPa$\cdot$mm$^{\text{1-m/2}}$$\cdot$cycles$^{-1}$\\[4pt]
 $m$  &&& 3.5   \\[4pt]
 $p$  & &&0.5   \\[4pt]
 $q$  &&& 1.0 \\[4pt]
 $K_c$  & = &$\sqrt{Y\gamma}$  & 64.8 MPa$\cdot$$\sqrt{\text{mm}}$ \\[4pt]
 $K_T$   &=  & 0.1$K_c$  &  6.5 MPa$\cdot$$\sqrt{\text{mm}}$  \\[4pt]
\toprule
\end{tabular}
\caption{Adopted parameters for the NASGRO equation.}
\label{tab:NASGRO_param}
\end{table}
	
	 The adopted geometry is the same presented in Fig.~\ref{fig:DCB}, however the fatigue tests are usually performed controlling the load instead of the displacement. Hence, the energy release rate function (\ref{eq:ERR_DCB_nondim_p}) is adopted hereafter to obtain the applied $\bs{\Delta} \bar G$. Constant amplitude load cycles featuring complete unloading are assumed for simplicity and, to ensure a pure opening mode, only tensile (i.e., positive) values of the load are applied\footnote{Under the assumption of perfectly brittle behavior, the extension to variable amplitude cycles and partial unloading is possible and straightforward. For materials whose fatigue behavior is sensitive to compression states, to the mean load value or to the presence of over- or under-loading events (i.e., for small scale plasticity) (\ref{eq:nasgro}) might be unrealistic. Rather, the introduction of plasticity and a data set dependent on \textit{ad-hoc} history variable are recommended as, e.~g., in \cite{Eggersmann2019}. However, the extensions to these behavior would substantially modify the derivation of the governing equations in sect.~\ref{sct:theory} and is out of the scope of the present paper.}.  Further assuming that the compliance of the specimen remains constant during a single cycle, the energy release rate excursion within the cycle $N_{k+1}$ can be obtained as

	\eqn{eq:D_ERR_Nk}{\bs{\Delta} \bar G_{k+1}= \bar G(\bar a_k, \bar P) = \frac{12\bar a_{k}^2}{\bar Y \bar b^2\bar h^3}\bar P^2\,,}

\noindent where $\bar P$ coincides with the maximum load applied.	

	Fig.~\ref{fig:comp_fat}b presents the comparison between reference and data-driven results in terms of crack growth curves for the noiseless and noisy data set for a maximum applied load $\bar P$ = 3.00, 2.50, 2.00 and 1.81. The value $\bar P$ = 1.81 is used since it is slightly higher than the theoretical fatigue-threshold load $\bar P_T$ = 1.80 obtained from (\ref{eq:D_ERR_Nk}) for $\bar a=\bar a_0$ and $\bs{\Delta} \bar G = \bar G_T$. Apart for the tests with $\bar P$ = 1.81, the data-driven results are always in excellent agreement with the reference curves regardless of the presence or not of noise. 
	
	Conversely, the presence of noise for $\bar P$ = 1.81 seems to affect the results, leading to an overestimation of the fatigue life by less than 15\%. This is due to the fact that for load levels close to the threshold value, the fatigue life is dominated by the sub-horizontal low crack-growth rate regime (sometimes referred to as nucleation or short-crack regime \cite{Newman1998}, Fig.~\ref{fig:comp_fat}a). In this phase the crack size and, hence, the applied energy release rate range grows very slowly leading to an initial sub horizontal branch of the crack growth curve characterized by a very low and almost constant crack-growth rate (Fig.~\ref{fig:comp_fat}b). Since for the fatigue sub-critical crack growth the data-driven search procedure aims at minimizing the difference between applied and resistant energy release rate, in this phase a relatively small perturbation in $\bar G_R$ might induce a significant variation in the initial crack growth rate that randomly leads to an over- or under-estimation of the fatigue life depending on the energy release rate applied and on the characteristics of the data set. However, it is well known and accepted in the fatigue community that experimental results performed in the short-crack regime are extremely subjected to the aleatory presence of imperfections at the microscale, that generate deviations from the expected crack evolution similar to what observed here \cite{Newman1998}.
	
	Fig.~\ref{fig:comp_fat}b shows also the results obtained with the noisy database of Fig.~\ref{fig:comp_fat}a adopting the cycle-jump technique illustrated in sect.~\ref{sct:num_fatigue} with ${\bs{\Delta}} N$ = 10. Although very simple, such approach allows to drastically reduce the computational time needed while preserving a good agreement with the reference results. Also in this case the test with $\bar P$ = 1.81 overestimates the fatigue life of about 15\%, further confirming that the highlighted mismatch is mainly due to the presence of noise. 
	
\begin{figure}[!h]
	\begin{adjustwidth}{-3cm}{-3cm}
	\centering
		\includegraphics{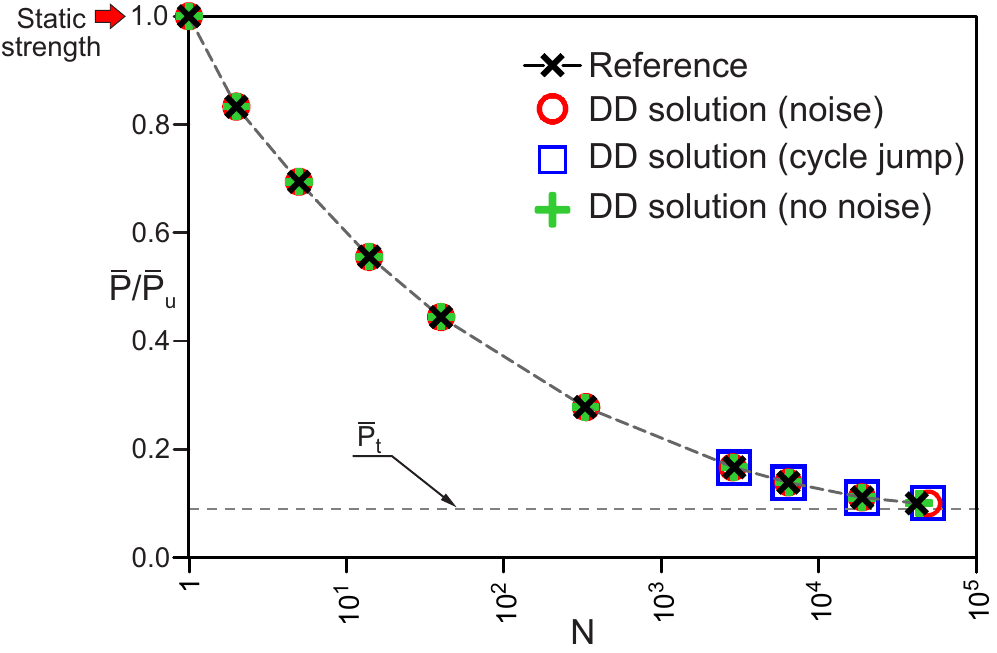}
	\end{adjustwidth}
		\caption{Comparison between reference and data-driven solutions in terms of modified W{\"o}hler curve.}
		\label{fig:wohler}
	\end{figure}

	Another important curve used to illustrate the effects of the fatigue loading is the $S-N$ or W{\"o}hler curve, which relates the applied load amplitude with the fatigue life in terms of maximum number of cycles that a specimen can sustain before failure. Fig.~\ref{fig:wohler} shows the comparison between the reference and the data-driven modified W{\"o}hler curves obtained. The agreement is again excellent along the whole range of load from the fatigue-threshold value to the static strength of the specimen $\bar P_u$, i.~e. the maximum load attainable in a monotonic quasi-static rate-independent test (Figs.~\ref{fig:Griff_comp}b,d). In particular, when such load is applied, the specimen fails as expected in a single cycle, thus bridging the critical and sub-critical crack growth. Also, the obtained curve has the typical shape of the W{\"o}hler curve that is obtained experimentally for brittle materials, with a tail that tends to flatten, further confirming the validity of the proposed approach.

	\FloatBarrier

\section{Summary and concluding remarks} \label{sct:conclusions}

We have extended the data-driven fracture mechanics approach to the case of rate-dependent and sub-critical fatigue  processes. As in \cite{Carrara2020}, the governing equations are derived from epistemic conservation laws in a variationally consistent way while we remove from the solution procedure any material modeling assumption characterizing the crack propagation. Rather, the fracture constitutive behavior is completely encoded into a discrete material data set. The solution of the crack propagation problem is determined as the point of the material data set that best fulfills a metastable or local stability principle. The latter condition is enforced by identifying the point in the material data set whose distance from the (analytically known) energy release rate function is minimum, following a closest-point-projection strategy. 

	For rate-dependent crack propagation two approaches are devised. The first deals with material data sets independent on the crack size, reproducing a fracture model with an implicit Griffith-type rate-independent limit. The second approach encompasses the more general case of material points dependent also on the crack size, enabling to reproduce also R-curve type rate-independent limit behaviors. The sub-critical fatigue crack growth process is tackled from the standpoint of the crack growth rate constitutive law, where the number of cycles plays the role of a pseudo-time and the crack growth is driven by the range of stress intensity factor or energy release rate spanned at the crack tip in a single cycle. A simple cycle jump approach is also presented to limit the computational time for high-cycle fatigue.
	
	The proposed approaches have been tested on double-cantilever-beam specimens with different geometries and using different artificially generated randomized material data sets, with and without noise, reproducing fatigue, Griffith and R-curve type rate-independent limit behaviors in brittle materials. After comparing the results with reference analytical solutions and with those obtained adopting the rate-independent approach \cite{Carrara2020}, the following conclusions can be drawn:
	
	\begin{itemize}
	\item[-] the proposed approaches delivers results in excellent agreement with those obtained solving the related fracture mechanics problems along classical analytical lines. In particular, the effect of the crack tip propagation velocity on the global response of the specimen is correctly reproduced and, for sufficiently low loading rates, the rate-dependent approach correctly reproduces the rate-independent results;
	 \item[-] the fatigue crack growth is correctly reproduced in terms of both crack growth and W{\"o}hler curves for both high- and low-cycle fatigue regimes. When the adopted database is affected by noise and the the behavior is dominated by the short-crack regime (i.~e., for applied loads close to the fatigue threshold) the procedure may over- or under-estimate the fatigue life, which however is consistent with what is typically observed experimentally;
	 \item[-] the robustness with respect to noisy data sets of the closest-point-projection strategy is confirmed;
	\item[-] the rate-dependent approach allows to follow and study the fast evolution of the system taking place in correspondence of the crack jumps in a rate-independent setting. This amends some pathological behaviors arising when adopting the rate-independent approach and multiple competing meta-stable states are present. In particular, it prevent the system from overcoming energetic barriers. In this sense, the proposed approach encompasses and extends the rate-independent data-driven fracture mechanics approach of \cite{Carrara2020};
	\item[-] the adoption of a data-driven rate-dependent fracture mechanics approach makes the introduction of some assumptions on the analytical evolutive relationships redundant, since they are implicitly encoded into the material data set. This happens, e.~g., with the irreversibility and the properties that the fracture constitutive law must have to render the problem convex.
	\end{itemize}

\FloatBarrier
\section*{Acknowledgements}
P. Carrara gratefully acknowledges the financial support of the German Research Foundation (DFG) through the Fellowship Grant CA~2359/1.

\bibliographystyle{unsrt}
\bibliography{Biblio}

\begin{thebibliography}{10}

\bibitem{Kirchdoerfer2016}
T~Kirchdoerfer and Michael Ortiz.
\newblock {Data-driven computational mechanics}.
\newblock {\em Computer Methods in Applied Mechanics and Engineering},
  304:81--101, jun 2016.

\bibitem{Conti2018}
S~Conti, S~M{\"{u}}ller, and Michael Ortiz.
\newblock {Data-Driven Problems in Elasticity}.
\newblock {\em Archive for Rational Mechanics and Analysis}, 229(1):79--123,
  jul 2018.

\bibitem{Lopez2018}
E.~Lopez, D.~Gonzalez, J.~V. Aguado, E.~Abisset-Chavanne, E.~Cueto,
  C.~Binetruy, and F.~Chinesta.
\newblock {A Manifold Learning Approach for Integrated Computational Materials
  Engineering}.
\newblock {\em Archives of Computational Methods in Engineering}, 25(1):59--68,
  jan 2018.

\bibitem{Carrara2020}
Pietro Carrara, L.~{De Lorenzis}, Laurent Stainier, and Michael Ortiz.
\newblock {Data-driven fracture mechanics}.
\newblock {\em Computer Methods in Applied Mechanics and Engineering},
  372:113390, dec 2020.

\bibitem{Kirchdoerfer2017}
T.~Kirchdoerfer and Michael Ortiz.
\newblock {Data Driven Computing with noisy material data sets}.
\newblock {\em Computer Methods in Applied Mechanics and Engineering},
  326:622--641, 2017.

\bibitem{Kirchdoerfer2018}
T.~Kirchdoerfer and Michael Ortiz.
\newblock {Data-driven computing in dynamics}.
\newblock {\em International Journal for Numerical Methods in Engineering},
  113(11):1697--1710, 2018.

\bibitem{Ibanez2017}
Ruben Iba{\~{n}}ez, Domenico Borzacchiello, Jose~Vicente Aguado, Emmanuelle
  Abisset-Chavanne, Elias Cueto, Pierre Ladev{\`{e}}ze, and Francisco Chinesta.
\newblock {Data-driven non-linear elasticity: constitutive manifold
  construction and problem discretization}.
\newblock {\em Computational Mechanics}, 60(5):813--826, nov 2017.

\bibitem{Kanno2019}
Yoshihiro Kanno.
\newblock {Mixed-integer programming formulation of a data-driven solver in
  computational elasticity}.
\newblock {\em Optimization Letters}, 13(7):1505--1514, 2019.

\bibitem{Eggersmann2020}
Robert Eggersmann, Laurent Stainier, Michael Ortiz, and Stefanie Reese.
\newblock {Model-free Data-Driven Computational Mechanics Enhanced by Tensor
  Voting}.
\newblock {\em arXiv:2004.02503v2}, pages 1--25.

\bibitem{Kanno2020}
Yoshihiro Kanno.
\newblock {A kernel method for learning constitutive relation in data-driven
  computational elasticity}.
\newblock {\em Japan Journal of Industrial and Applied Mathematics}, 2020.

\bibitem{Conti2020}
S.~Conti, S.~M{\"{u}}ller, and Michael Ortiz.
\newblock {Data-Driven Finite Elasticity}.
\newblock {\em Archive for Rational Mechanics and Analysis}, 237(1):1--33, jul
  2020.

\bibitem{Nguyen2018}
Lu~Trong~Khiem Nguyen and Marc~Andr{\'{e}} Keip.
\newblock {A data-driven approach to nonlinear elasticity}.
\newblock {\em Computers and Structures}, 194:97--115, 2018.

\bibitem{Leygue2018}
Adrien Leygue, Michel Coret, Julien R{\'{e}}thor{\'{e}}, Laurent Stainier, and
  Erwan Verron.
\newblock {Data-based derivation of material response}.
\newblock {\em Computer Methods in Applied Mechanics and Engineering},
  331:184--196, 2018.

\bibitem{Stainier2019}
Laurent Stainier, Adrien Leygue, and Michael Ortiz.
\newblock {Model-free data-driven methods in mechanics: material data
  identification and solvers}.
\newblock {\em Computational Mechanics}, 64(2):381--393, aug 2019.

\bibitem{Flaschel2020}
Moritz Flaschel, Siddhant Kumar, and Laura {De Lorenzis}.
\newblock {Unsupervised discovery of interpretable hyperelastic constitutive
  laws}.
\newblock 2020.

\bibitem{Eggersmann2019}
R.~Eggersmann, T.~Kirchdoerfer, S.~Reese, Laurent Stainier, and Michael Ortiz.
\newblock {Model-Free Data-Driven inelasticity}.
\newblock {\em Computer Methods in Applied Mechanics and Engineering},
  350:81--99, jun 2019.

\bibitem{Ladeveze2019}
Pierre Ladev{\`{e}}ze, David N{\'{e}}ron, and Paul~William Gerbaud.
\newblock {Data-driven computation for history-dependent materials}.
\newblock {\em Comptes Rendus - Mecanique}, 347(11):831--844, 2019.

\bibitem{Lefranc2014}
Maxime Lefranc and Elisabeth Bouchaud.
\newblock {Mode I fracture of a biopolymer gel: Rate-dependent dissipation and
  large deformations disentangled}.
\newblock {\em Extreme Mechanics Letters}, 1(2014):97--103, 2014.

\bibitem{Hauch1998}
J.~A. Hauch and M.~P. Marder.
\newblock {Energy balance in dynamic fracture, investigated by a potential drop
  technique}.
\newblock {\em International Journal of Fracture}, 90(1-2):133--151, 1998.

\bibitem{Negri2010b}
Matteo Negri.
\newblock {From Rate-Dependent to Rate-Independent Brittle Crack Propagation}.
\newblock {\em Journal of Elasticity}, 98(2):159--187, feb 2010.

\bibitem{Negri2010}
Matteo Negri.
\newblock {A comparative analysis on variational models for quasi-static
  brittle crack propagation}.
\newblock {\em Advances in Calculus of Variations}, 3(2):149--212, 2010.

\bibitem{Toader2009}
Rodica Toader and Chiara Zanini.
\newblock {An artificial viscosity approach to quasistatic crack growth}.
\newblock {\em Bolletino dell Unione Matematica Italiana}, 2(1):1--35, 2009.

\bibitem{Knees2008}
Dorothee Knees, Alexander Mielke, and Chiara Zanini.
\newblock {Model for Crack Propagation}.
\newblock 18(9):1529--1569, 2008.

\bibitem{Larsen2009}
C.~J. Larsen, Michael Ortiz, and C.~L. Richardson.
\newblock {Fracture Paths from Front Kinetics: Relaxation and Rate
  Independence}.
\newblock {\em Archive for Rational Mechanics and Analysis}, 193(3):539--583,
  sep 2009.

\bibitem{Fineberg1999}
Jay Fineberg, Steven~P Gross, M~Marder, and Harry~L Swinney.
\newblock {Fast Cracks}.
\newblock {\em Science}, 284(5418):1233d--1233, 1999.

\bibitem{Sharon1999}
Eran Sharon and Jay Fineberg.
\newblock {Confirming the continuum theory of dynamic brittle fracture for fast
  cracks}.
\newblock {\em Nature}, 397(6717):333--335, 1999.

\bibitem{Ravi2004}
K.~Ravi-Chandar.
\newblock {Dynamic Fracture}.
\newblock {\em Dynamic Fracture}, pages 1--254, 2004.

\bibitem{Paris1963}
Paul~C. Paris and F~Erdogan.
\newblock {A critical analysis of crack propagation laws}.
\newblock {\em Journal of Basic Engineering}, 85(4):528--533, 1963.

\bibitem{Rabold2013}
Frank Rabold, Meinhard Kuna, and Thomas Leibelt.
\newblock {Procrack: A software for simulating three-dimensional fatigue crack
  growth}.
\newblock {\em Lecture Notes in Applied and Computational Mechanics},
  66:355--374, 2013.

\bibitem{Cojocaru2006}
D.~Cojocaru and A.~M. Karlsson.
\newblock {A simple numerical method of cycle jumps for cyclically loaded
  structures}.
\newblock {\em International Journal of Fatigue}, 28(12):1677--1689, 2006.

\bibitem{Oskay2004}
Caglar Oskay and Jacob Fish.
\newblock {Fatigue life prediction using 2-scale temporal asymptotic
  homogenization}.
\newblock {\em International Journal for Numerical Methods in Engineering},
  61(3):329--359, 2004.

\bibitem{Bhattacharyya2019}
Mainak Bhattacharyya, Am{\'{e}}lie Fau, Rodrigue Desmorat, S.~Alameddin,
  D.~N{\'{e}}ron, Pierre Ladev{\`{e}}ze, and Udo Nackenhorst.
\newblock {A kinetic two-scale damage model for high-cycle fatigue simulation
  using multi-temporal Latin framework}.
\newblock {\em European Journal of Mechanics, A/Solids}, 77(October
  2018):103808, 2019.

\bibitem{Hutchinson1979}
J.~W. Hutchinson.
\newblock {\em {A course in Nonlinear Fracture Mechanics}}.
\newblock Technical University of Danmark, Lyngby, 1979.

\bibitem{Alessi2018}
Roberto Alessi, Vito Crismale, and Gianluca Orlando.
\newblock {Fatigue effects in elastic materials with variational damage models:
  A vanishing viscosity approach}.
\newblock pages 1--30, jul 2018.

\bibitem{Newman1998}
J.C. Newman.
\newblock {The merging of fatigue and fracture mechanics concepts: a historical
  perspective}.
\newblock {\em Progress in Aerospace Sciences}, 34(5-6):347--390, 1998.

\end{thebibliography}

\appendix

\section{Data-driven search algorithms} \label{app:algorithms}

The pseudo-code for the implementation of the data-driven search procedures are detailed in the following.

	\subsection{Implicit quasi-static Griffith model} \label{app:griffith}
	
	The data-driven search algorithm for the implicit quasi-static Griffith model is presented in Algorithm~\ref{algo:DD_Griff}.

\begin{algorithm}[!ht]
	 \DontPrintSemicolon
	\setstretch{1.2}
	$\bs{Step}:\,\,k+1$

	\KwIn{$\,\,\Delta_{T,k+1}=\Delta_{T,\,t_k+\bs{\Delta} t}$, $a_{k}\,,v_k,\, \hat G_{R}^{QS}$}

	\KwOut{{$v_{k+1},\,a_{k+1},\,G_{R\,\,k+1},\,\,\Delta_{k+1},\,P_{k+1},\,\,G_{DD,k+1}$}}

	\tcc{BEGINNING OF THE COMPUTATION}

	\bf{Define}: $\bs{\Delta} a_{k+1}^{TRIAL} = \displaystyle\frac{v_k}{2}\bs{\Delta} t$\;
	
	\tcc{Compute the solution}

	\uIf{$G(\Delta_{T,k+1},a_k+\bs{\Delta} a_{k+1}^{TRIAL})-\hat G_R^{QS}< 0$}{\vspace{-5mm}\tcp*{Crack arrest}
	
	\bf{Assign}: $v_{k+1} = 0$
	\begin{adjustwidth}{0cm}{0cm}
	\hspace{16mm}$\bs{\Delta} a_{k+1}=\bs{\Delta} a_{k+1}^{TRIAL}$
	
	 \hspace{16mm}$G_{DD,k+1}=G(\Delta_{T,k+1},a_k+\bs{\Delta} a_{k+1})$\;
	\end{adjustwidth}}
	
	\Else{\vspace{-5mm}\tcp*{Crack propagation}
	
	{\bf{Compute}:
	\begin{adjustwidth}{0cm}{-2cm}
	$d_{i,\,k+1}= {\rm \underset{\mathit{ a \ge a_k}}{min}} \left\{\sqrt{\left(\displaystyle\frac{v_k+\hat v_i}{2}\bs{\Delta}t+a_k- a\right)^2+\left(\hat G_{R,i} - G(\Delta_{T,k+1}, a)\right)^2}\right\}$

        \vspace{3mm}$i^*_{k+1}= {\rm \underset{\mathit{i}}{arg\,min}}\left\{d_{i,\,k+1}\ :\ (\hat v_{i},\,\hat G_{R,i})\in\mathcal{D}_{R}\right\}$\;
	\end{adjustwidth}}
	
       { \vspace{3mm}\bf{Assign}: $v_{k+1} = \hat v_{i^*_{k+1}}$
	\begin{adjustwidth}{0cm}{0cm}
	\hspace{16mm}$\bs{\Delta} a_{k+1}=\displaystyle\frac{v_k+\hat v_{i^*_{k+1}}}{2}\bs{\Delta} t$\;
	
	 \vspace{2mm}\hspace{16mm}$G_{DD,k+1}=\hat G_{R,i^*_{k+1}}$\;
	\end{adjustwidth}}

}
		\bf{Compute}: $a_{k+1} = a_k + \bs{\Delta} a_{k+1}$
		\begin{adjustwidth}{0cm}{-2cm}
		\vspace{2mm}\hspace{20mm}$\left(\Delta_{k+1},\,P_{k+1}\right) = \left(\displaystyle\frac{C\left(a_{k+1}\right)}{C\left(a_{k+1}\right)+C_{M}}\Delta_{T,k+1},\, \displaystyle\frac{\Delta_{T,k+1}}{C\left(a_{k+1}\right)+C_{M}}\right)$ 
		\end{adjustwidth}
		
	\caption{Data-driven fracture mechanics algorithm - Implicit quasi-static Griffith model. Given: ${G}(\Delta_T,\, a)$, $\,\,C(a)$, $\,\,\mathcal{D}_R$, $\,\,\Delta_{T,\,t}$.}
	\label{algo:DD_Griff}
\end{algorithm}

	\subsection{Complete model} \label{app:complete}

The data-driven search algorithm for the complete model is presented in Algorithm~\ref{algo:DD_3D}.	

\begin{algorithm}[!hb]
	 \DontPrintSemicolon
	\setstretch{1.2}
	$\bs{Step}:\,\,k+1$

	\KwIn{$\,\,\Delta_{T\,k+1}=\Delta_{T\, t_k+\bs{\Delta} t}$, $a_{k}\,,v_k,\, \hat G_{R}^{QS}$}

	\KwOut{{$v_{k+1},\,a_{k+1},\,G_{R\,\,k+1},\,\,\Delta_{k+1},\,P_{k+1},\,\,G_{DD,k+1}$}}

	\tcc{BEGINNING OF THE COMPUTATION}
	
	\bf{Set}: $\bs{\Delta} a(v) = \displaystyle \frac{v_k+v}{2}\bs{\Delta} t$

	\bf{Define}: $\bs{\Delta} a_{k+1}^{TRIAL} = \displaystyle\frac{v_k}{2}\bs{\Delta} t$

	\bf{Compute}: $\hat G_{R,k+1}^{QS}=\left\langle \hat G_{R,i_L}^{QS},\,\hat G_{R,i_S}^{QS}\right\rangle$\;
	
	\tcc{Compute the solution}

	\uIf{$G(\Delta_{T,k+1},a_k+\Delta a_{k+1}^{TRIAL})-\hat G_{R,k+1}^{QS}< 0$}{\vspace{-5mm}\tcp*{Crack arrest}
	
	\bf{Assign}: $v_{k+1} = 0$
	\begin{adjustwidth}{0cm}{0cm}
	\hspace{16mm}$\bs{\Delta} a_{k+1}=\bs{\Delta} a_{k+1}^{TRIAL}$
	
	 \hspace{16mm}$G_{DD,k+1}=G(\Delta_{T,k+1},a_k+\bs{\Delta} a_{k+1})$\;
	\end{adjustwidth}
	}
	
	\Else{\vspace{-5mm}\tcp*{Crack propagation}

	\bf{Compute}:
	\begin{adjustwidth}{0cm}{-2cm}
	$\begin{array}{ll}&d_{i,\,k+1}={\rm \underset{\mathit{v\ge0}}{min}} \left\{\ \vphantom{\sqrt{\left(\hat G_{R,i}\right)}}\right.\\ &\left. \sqrt{\left(\hat v_{i}-v\right)^2+\left(\hat a_i-a_k-\bs{\Delta} a(v)\right)^2+\left(\hat G_{R,i} - G(\Delta_{T,k+1},a_k+\bs{\Delta} a(v)\right)^2}\right\}\end{array}$

        \vspace{3mm}\hspace{5mm}$i^*_{k+1}= {\rm \underset{\mathit{i}}{arg\,min}}\left\{d_{i,\,k+1}\ :\ (\hat v_i,\,\hat a_i, \,\hat G_{R,i})\in \mathcal{D}_R\right\}$\;
	\end{adjustwidth} 
	
	\nl\uIf{$\hat v_{i^*}\notin\mathcal{D}_R^{QS}$ }{ \vspace{-6mm}\tcp*{Rate-dep. propagation}\vspace{1mm}	

	\bf{Assign}: $v_{k+1} = \hat v_{i^*_{k+1}}$\; 
	
	\hspace{16mm}$\bs{\Delta} a_{k+1}=\displaystyle\frac{v_k+v_{i^*_{k+1}}}{2}\bs{\Delta} t$\;
	
	 \hspace{16mm}$G_{DD,k+1}=\hat G_{R,i^*_{k+1}}$\;}

	\nl\uElse{ \vspace{-6mm}\tcp*{Quasi-static propagation}\vspace{1mm}	\bf{Assign}: $v_{k+1} = 0$\; 

	\hspace{16mm}$\bs{\Delta} a_{k+1}=\hat a_{i^*_{k+1}}-a_k$\;
	
	 \hspace{16mm}$G_{DD,k+1}=\hat G_{R,i^*_{k+1}}$
	 }
	 \nl\bf{end}
	}
	
	\bf{Compute}: $a_{k+1} = a_k + \bs{\Delta} a_{k+1}$
		\begin{adjustwidth}{0cm}{-2cm}
		\vspace{2mm}\hspace{20mm}$\left(\Delta_{k+1},\,P_{k+1}\right) = \left(\displaystyle\frac{C\left(a_{k+1}\right)}{C\left(a_{k+1}\right)+C_{M}}\Delta_{T,k+1},\, \displaystyle\frac{\Delta_{T,k+1}}{C\left(a_{k+1}\right)+C_{M}}\right)$ 
		\end{adjustwidth}

	 \caption{Data-driven fracture mechanics algorithm - Complete model. Given: ${G}(\Delta_T,\, a)$, $\,\,C(a)$, $\,\,\mathcal{D}_R$, $\,\,\Delta_{T,\,t}$.}\label{algo:DD_3D}
\end{algorithm}	

	\subsection{Sub-critical crack growth and fatigue} \label{app:fatigue}
	
	The data-driven search algorithm for the sub-critical and fatigue crack growth is presented in Algorithm~\ref{algo:DD_fat}.
	
	\setlength{\algotitleheightrule}{\algotitleheightruledefault}
	\begin{algorithm}[!ht]
	 \DontPrintSemicolon
	\setstretch{1.2}
	$\bs{Step}:\,\,k+1$

	\KwIn{$\,\,a_{k}\,,N_k,\, \hat G_{f}^{T},\,\bs{\Delta} N $}

	\KwOut{{$v_{k+1},\,a_{k+1},\,G_{DD,k+1},\,\bs{\Delta} \Delta_{k+1}\ \text{and/or}\ \bs{\Delta} P_{k+1},\,N_{k+1}$}}

	\tcc{BEGINNING OF THE COMPUTATION}

\begin{adjustwidth}{0cm}{-2cm}
	\bf{Define}: $\bs{\Delta} G_{k+1} = G(\bullet_{max},a_k)- G(\bullet_{min},a_k) $\;
	\end{adjustwidth}
	
	\tcc{Compute the solution}

	\uIf{$\bs{\Delta} G_{k+1} - \hat G_f^T < 0$}{\vspace{-5.8mm}\tcp*{Infinite fatigue life}
	
	\vspace{1mm}\bf{No propagation $\rightarrow$ EXIT}}
	
	\Else{\vspace{-6.7mm}\tcp*{Crack propagation}
	
	{\bf{Compute}:
	\begin{adjustwidth}{0cm}{-2cm}
        \vspace{3mm}$i^*_{k+1}= {\rm \underset{\mathit{i}}{arg\,min}}\left\{\bs{\Delta} G_{k+1}-\hat G_{f,i}\ :\ \hat G_{f,i}\in\mathcal{D}_{R,f}\right\}$\;
	\end{adjustwidth}}
	
       { \vspace{3mm}\bf{Assign}: $v_{k+1} = \hat v_{i^*}$
	\begin{adjustwidth}{0cm}{0cm}
	\hspace{16mm}$\bs{\Delta} a_{k+1}=\bs{\Delta} N\cdot\hat v_{i^*}$
	
	\hspace{16mm}$\bs{\Delta} G_{DD,k+1}=\hat G_{f,i^*}$\;
	\end{adjustwidth}}

}
		\bf{Compute}: $a_{k+1} = a_k + \bs{\Delta} a_{k+1}$
		\begin{adjustwidth}{0cm}{-2cm}
		\vspace{2mm}\hspace{20mm}$N_{k+1}=N_k+\bs{\Delta} N$
		
		\vspace{2mm}\hspace{20mm}$\bs{\Delta} P_{k+1} = \displaystyle\frac{\Delta_{max}-\Delta_{min}}{C(a_{k+1})} \quad\quad \text{\normalfont \textsf{($\Delta$-control)}}$ 

		\vspace{2mm}\hspace{20mm}$\bs{\Delta} \Delta_{k+1} = \left(P_{max}-P_{min}\right) C(a_{k+1}) \quad\quad \text{\normalfont \textsf{($P$-control)}}$ 
		
		\begin{adjustwidth}{0cm}{-2cm}
		\vspace{2mm}\hspace{16.5mm}$\begin{array}{ll}\left(\bs{\Delta} \Delta_{k+1},\,\bs{\Delta} P_{k+1}\right) = &\left(\displaystyle\frac{C\left(a_{k+1}\right)}{C\left(a_{k+1}\right)+C_{M}}(\Delta_{T,max}-\Delta_{T,min}),\,\right. \\[15pt] &\left.  \displaystyle\frac{\Delta_{T,max}-\Delta_{T,min}}{C\left(a_{k+1}\right)+C_{M}}\right)\quad\quad\text{\normalfont \textsf{($\Delta_T$-control)}} \end{array}$ 
		\end{adjustwidth}
		\end{adjustwidth}
		
	\caption{Data-driven fracture mechanics algorithm - Sub-critical fatigue crack growth. Given: ${G}(\bullet,\, a)$, $\,\,C(a)$, $\,\,\mathcal{D}_{R,f}$, $\,\,\bullet_{max}$, $\,\,\bullet_{min}$.}
	\label{algo:DD_fat}
\end{algorithm}

\FloatBarrier

\end{document}